%% file: main.tex
\documentclass{article}
\PassOptionsToPackage{numbers,sort&compress}{natbib}
\usepackage[preprint]{neurips_2026_vericode}
\usepackage[utf8]{inputenc}
\usepackage[T1]{fontenc}
\usepackage{url}
\usepackage{booktabs}
\usepackage{float}
\usepackage{flafter}
\usepackage{hyperref}
\usepackage{amsmath,amsfonts}
\usepackage{microtype}
\usepackage{graphicx}
\usepackage{xcolor}
\hypersetup{pdftitle={From Experiments to Decisions: Reusing Evidence in Autonomous Coding Research},pdfauthor={Bobber Cheng},hidelinks}

\title{From Experiments to Decisions:\\Reusing Evidence in Autonomous Coding Research}
\author{Bobber Cheng}

\begin{document}
\maketitle

\input{body}

\label{bodyend}
\clearpage
\bibliographystyle{plainnat}
\bibliography{references}

\clearpage
\appendix

\noindent\textbf{Appendix reading guide.}
Appendix~\ref{app:decisions} exposes the decision and process evidence;
Appendices~\ref{app:numericaldetail}--\ref{app:witness} unpack the numerical result;
Appendix~\ref{app:spatial} preserves the spatial and execution-gate companion;
\ref{app:provenance}--\ref{app:screenbounds} expose historical selection, linkage and negative branches;
\ref{app:replay}--\ref{app:environment} describe execution and provenance limits.
Campaign, resource, effort and cross-domain context follows in
Appendices~\ref{app:campaign}--\ref{app:crossdomain}.

\input{decision_appendix}

\input{numerical_details}

\section{A source-grounded arithmetic model of P002}
\label{app:schedule}

This is an explanation of the disclosed graph under an explicit model, not a new discovery run or native execution. We inspected Microsoft's immutable ONNX Runtime source revision \texttt{2d924974ef147392ced8409d36bd6d2e7fcc8a74}, which resolves its 1.24.4 tag~\citep{orteinsum}. The supplementary replay guide supplies file-level hashes and locations for the entry point, preprocessing, typed processor, CPU helpers, and matrix-multiplication dispatch.

\paragraph{From source to recurrence.}
The CPU entry point collects operands in input-index order. Preprocessing rearranges axes within operands without changing that order. The typed processor first reduces labels occurring only in input zero, then pairs the previous result with each next input. Each pair writes a newly allocated float-typed tensor before that result is consumed again. Applied to P002, the first 23 operands independently reduce to color counts; operand 23 (zero-based) reduces to occupancy. The 26 scalar operands follow separately, not as a precomputed power. The derived pairwise matrix shapes $(\text{batch},M,K,N)$ are $(10,1,1,1)$ for additional count factors, $(1,10,1,900)$ for occupancy, and $(1,9000,1,1)$ for scalars. In each pair $K=1$, so there is no sum of multiple nontrivial products. Earlier 0/1 reductions have exact small integer results for these cases.

Under correctly rounded nearest-even binary32 multiplication with gradual underflow, this yields the recurrence in Appendix~\ref{app:numericaldetail}. For count two, the last five rounded values in units of $u$ are
\[
51404300,\quad510496,\quad5070,\quad50,\quad0.
\]
The penultimate exact product is $5070a=6757881585/134217728$ units, which rounds to 50. The final product is $50a=66645775/134217728$ units, strictly below one half; zero follows. Count three ends at 5611 units and count one at zero. The standard-library reference decodes binary32 bit fields and rounds exact rational products, with explicit tie handling across normal and subnormal boundaries. It never calls the native runtime or uses historical output amplitudes as operands.

\paragraph{What the match does and does not establish.}
This is a post-hoc investigation: the archived final outputs were already known. We inspected one source-determined order and used the unchanged disclosed scalar, not a search across associations or coefficients. The calculation matches the historical amplitudes and explains a concrete intermediate-rounding boundary under its assumptions. However, the inspected 1.24.4 source corresponds to the September replay's reported version, not July's environment linked to 1.27.0. The replay wheel's exact build-to-source identity, selected architecture kernel, floating-point controls, and historical intermediate values remain unverified. A changed contraction order or subnormal-flush policy can invalidate the calculation. Numerical agreement is therefore explanatory consistency, not native-path authentication, held-out prediction, or evidence that this diagnosis caused the original discovery.

\section{Numerical witness: specification, proof, and controls}
\label{app:witness}

Equation~\ref{eq:witness} is a newly constructed diagnostic, not a reconstruction of the historical ONNX graph. Its input contract is a fully occupied $3\times3$ grid with ten possible color channels, one count equal to three, all other counts at most two, and zero output outside the occupied region. The retained default generator produces the nonzero count profiles $(3,2,2,1,1)$ and $(3,2,1,1,1,1)$; direct integer-grid checks find 195 and 70 stored cases respectively. The supplement supplies all 265 de-identified rows in \path{task129_count_profiles.csv}. Run \texttt{python3 task129\_contract.py -{}-check task129\_count\_profiles.csv} to recompute their shape, frequency-gap, and mode-fill summaries; source holders can add \texttt{-{}-source TASK\_JSON} for read-only re-extraction without running a generator or model. The static source is ARC-GEN's \path{tasks/task_5582e5ca.py} at retained revision \texttt{a15cbdb44c77}, default size-three branch. Its multiplicities arise from six distinct colors used three, two, one, one, one, and one times, optionally merging the last two. The proved mathematical contract, derived retained-corpus audit, and unrecovered private-platform domain remain separate; neither the table nor the static reading authenticates hidden inputs.

For nonnegative exact values near zero, nearest-even binary32 rounding with subnormals retained maps values at most $u/2$ to zero and values strictly above $u/2$ to a positive value. Thus the coefficient $k$ in $n_c^2k$ separates counts one and two from three precisely when
\begin{equation}
\label{eq:separator}
9k>u/2\quad\text{and}\quad4k\leq u/2,
\qquad\text{equivalently}\qquad u/18<k\leq u/8.
\end{equation}
The lower endpoint fails because the tie rounds to even zero; the upper endpoint is allowed for the same reason. The chosen $k=u/16$ satisfies Equation~\ref{eq:separator}. Count zero and zero occupancy give exact zero. On the stated count contract there is consequently exactly one positive color at every occupied position and none on padding, which is the required mode fill. Unscaled $n_c^2$ has identical exact-positive support but a different rounded/decoded output whenever another color is present. This supplies the counterexample to treating that support alone as sufficient; no new general result in abstract interpretation is claimed.

Run \texttt{python3 mechanism\_witness.py} from the supplement with standard-library Python. It implements exact rational nearest-even rounding and independently checks the tested values using binary32 pack/unpack conversion. Their intermediate binary64 representations are exact for these dyadic inputs. For a synthetic grid with profile $(3,2,2,1,1)$, the script constructs and checks all 9,000 Boolean entries of the padded $(1,10,30,30)$ decoded output against an independently constructed mode-fill target. The round-once witness matches. Exact, binary64, and unscaled controls each produce 36 extra positive entries; early coefficient rounding and flush-to-zero each miss the nine required positives. These are controlled changes to this new demonstration, not historical candidate ablations.

The witness also checks all 15,540 labeled count signatures under the stated count contract; 13,860 have either retained default-generator count profile. These are finite algebraic signatures, not additional task cases, full-program evaluations, independent discoveries, or probability-weighted generator samples. Counts $(4,3,1,1)$ deliberately violate the contract and produce two selected colors: this is a threshold separator, not a general argmax. The exact expression, input domain, rounding location, and subnormal handling are all necessary parts of the demonstrated claim. The original candidate uses different parameters and runtime-dependent operations, whose correctness and cost remain supported by its separate historical receipts and archived-program replay.

\input{spatial_appendix}

\section{Historical selection, provenance, and verification boundaries}
\label{app:provenance}

\begin{table}[H]
\caption{Selection and coverage of the derived evidence. Completeness is asserted only within the stated retained unit, not across all historical research. Filenames refer to the supplement's \texttt{data/} directory.}
\centering
\begin{tabular}{@{}p{0.27\linewidth}p{0.28\linewidth}p{0.37\linewidth}@{}}
\toprule
File / unit & Selection rule & Coverage and excluded work \\
\midrule
\path{competition_progress.csv} & Exact initial and deadline archives & All 400 tasks; 394 with acceptance records at both endpoints; 381 after excluding mixed success flags \\
\path{tasks.csv} & Archived perfect set by endpoint & All 9 tasks; 3 competition-phase, 6 later additions; not all possible perfect tasks \\
\path{task053_candidates.csv} & Primary rows in cycles 1--9 of one retained run & All 105 rows; rechecks, other workers, earlier work and integration cycle excluded \\
\path{original_runtime_attempts.csv} & Curated returned recipient results & 7 milestones; not a complete original-attempt census \\
\path{original_semantic_screen.csv} & Retained eight-task screen & All 8 task rows; historical choice of this task subset is unrecovered \\
\path{fixed_artifact_recipients.csv} & Two already-successful fixed artifacts & All 800 recipient dispositions; no new-task holdout or family-eligibility cohort \\
\path{bounded_validations.csv} & Final post-reset state of one lane & All 264 rows in that state; earlier smoke/pilot omitted \\
\path{task129_events.csv} & Bounded closing-episode milestones & Public-target information through promotion; earlier routes incomplete \\
\path{input_features.csv} & Scoped interactive operator records & 483 included plus 5 barriers; missing responses and off-client preparation remain unknown \\
\bottomrule
\end{tabular}
\end{table}

The eight semantic-screen tasks are the subset named in the retained worker result; why that historical subset was chosen is unrecovered. The seven runtime results are a retrospectively curated set of directly returned failure/success milestones corroborated by later validation or bank records, not an algorithm-defined or exhaustive sample. Their denominators cannot support recipient success probabilities. Generator revisions and random seeds are not uniformly recovered, so the stated sample counts must not be treated as a homogeneous held-out test set.

The supplement's legacy \texttt{inherited} label groups three selected programs with public-comparison ancestry. It does not establish when the operator discovered their solutions or exclude locally constructed equivalents. The local gate admits only strictly cheaper replacements; at cost zero, even a correct local reconstruction need not replace the incumbent. First admission, local construction, discovery, and final artifact selection are therefore separate events.

The source audit distinguishes returned validation from test launch and bank promotion. For Task 053, the examined dispatch-to-launch interval is 401.934 seconds and dispatch-to-return is 438.484 seconds. The successful computation occurs inside the batch; its exact first internal timestamp is not retained. The nine-cycle run's last cycle overlaps this worker's dispatch. Its subsequent cycle imports a prevalidated artifact, so including that cycle as a fresh independent success would change the scientific unit.

For Task 129, the earliest perfect official-validation receipt identified within the bounded closing episode precedes the stronger validation by 121.496 seconds; the stronger receipt precedes ledger promotion by 71.725 seconds. Public target information precedes the no-answer-inspection request. Neither the first identified receipt nor the instruction establishes globally first or contamination-free invention. Selected failed batches are milestones, not a census of every attempted route.

The broad F1 lane has a smoke test and an initial 256-relation-per-task pilot before a reset. Its 264 retained post-reset validations exclude those earlier checks. The sum of their recorded isolated durations is 61.5236 seconds, while the reset checkpoint's creation-to-update interval is 588 seconds. Neither is aggregate CPU usage, human time, complete lane duration, or a substitute for missing model consumption. Candidate generation, semantic screening, and concurrent execution cannot be charged by summing these unlike measurements.

The supplement separates task costs and acceptance flags, candidate outcomes, recipient screens, relative-time events, effort features, and the mechanism evidence packet. Rejected candidates have explicitly unestablished correctness and blank semantic counts. Its verifier recomputes competition cohorts/gains, candidate/cycle and screen partitions, perfect-task correctness/cost predicates, event chronology and linkage consistency, replay-summary totals, and conditional effort arithmetic. The standalone witness executes the separate numerical diagnostic. Private source checks bind selected excerpts and links to retained records; the anonymous package cannot independently authenticate them. Neither de-identification nor a passing check certifies the interpretation of the history.

\section{Mechanism-trace observations and interpretation}
\label{app:mechanismtrace}

Table~\ref{tab:trace} groups retained observations by the inference at issue. In the supplement, \path{mechanism_evidence.json} links these groups to claims K001--K009, opaque events, candidate versions, and call/return or order-only links; \path{MECHANISM_EVIDENCE.md} explains how to inspect them. Excerpt X001 is the prior exclusion returned in M014; X002 is the numerical explanation in M018. These are short contiguous source excerpts with whitespace normalized, not newly authored quotations. Other event descriptions are labeled analyst paraphrases, and interpretations are separately marked. Private audits check source hashes and original linkage, but public consistency checks do not authenticate the underlying history or establish cognitive causation. Full conversations and original model files remain unbundled; P002's construction metadata is exposed separately.

\begin{table}[h]
\caption{History index: observed results and the records limiting their interpretation. Rows group retained units, not independent trials or matched controls; detailed partitions follow.}
\label{tab:trace}
\centering
\begin{tabular}{@{}p{0.17\linewidth}p{0.40\linewidth}p{0.35\linewidth}@{}}
\toprule
Retained unit & Recorded observations & Scope-limiting evidence \\
\midrule
F2 closing episode (K001--K003, K008) & X001/M014 exclusion; X002/M018 explanation. P001/M021 and tuned P002/M023 each pass 265/265 at cost one; only P002 adds 20,000 custom cases. & M015 is an implementation failure; M016--M017 are unsuccessful routes. Probe N002 tests an out-of-contract count-two mode. No complete attempt census or isolated causal change. \\
F1 gates (K004--K007) & 105 primary candidates; separate worker batches of 33, 24 and 88 records. The last has 47 configurations passing 60/60 at cost zero and 41 checker rejects. & Rejected execution leaves correctness unobserved. M006 consolidates memory after success. Do not pool the batches or infer a retrieval-to-gain chain. \\
F1 adaptation lineage (K009) & G001 fails the 087/140 corpora; G002 passes them. P003 separately links a recorded validation receipt to final bytes. The 385 mirror branch passes 0/265. & Each rotation construction serves both tasks. Early byte identities are unknown; G002 is not established as P003. Neither independent repairs nor general task inapplicability follows. \\
\bottomrule
\end{tabular}
\end{table}

For F2, the verified prior exclusion is returned at 02:40:42.467 UTC on July 18. Further audit material is retrieved later, but we do not assert it repeats that exclusion. The numerical explanation is recorded at 02:58:47.965, candidate creation at 02:58:47.968, test launch at 02:58:58.398, and the linked official receipt at 02:58:58.779. These event times identify ordering, not the moment of conception or a twelve-second discovery process. The later 03:01:00.275 receipt concerns a tuned candidate with different bytes, which independently passes the 265 official cases and 20,000 custom stress cases. The earlier traceback and two negative batches are route failures; we do not assert that each falsifies the support abstraction or that one caused the next hypothesis. The report's six-cell residual motivates a distinction between mode selection and complete output filling, but its causal role is interpretation.

\paragraph{The actual searched grammar and the separate scope argument.}
For binary one-hot predicates $x$, the earlier relational expressions have the form $T_q(x)[c,h,w]=\sum_{\mathbf z}\prod_{j=1}^{k}x[a_j(c,h,w,\mathbf z)]$, with typed spatial/color indices, connected positive-conjunctive cores, and a practical six-atom cap. They reuse input operands without an initializer. The implementation evaluates up to two authored and two generated probes using float32 NumPy contractions with greedy planning, thresholds above zero, and packs the Boolean outputs into a signature. A bounded beam retains up to three representatives per signature, alongside projection, positive-cover, and contraction-work filters. A matching signature is not equivalence over all inputs. Probe-exact survivors proceed to full-canvas checking. Selected literal source excerpts and static/source-holder checks in \path{history/} expose the numerical operations, repeated-input builder and recorded defaults without replaying or authenticating the original search.

Excerpt X001's scalar-irrelevance claim is a separate support argument for not expanding that grammar, not the implementation of the numerical signature. The later construction adds a scalar and exceeds the earlier atom bound. Neither P001 nor P002 is shown to have been generated and discarded by signature deduplication. The contemporary explanation challenges that rationale; it does not isolate scalar addition, higher arity, or changed execution order as the sole beneficial intervention. The new witness supplies a falsifier for the unrestricted support argument, not a historical pruning-error proof.

\paragraph{Historical numerical observations, distinct from the new witness.}
Table~\ref{tab:historicalprobes} exports designated-channel probes from P002's contemporary stronger receipt M023. A probe uses a full nine-cell input with one color occurring $n$ times and every remaining cell a distinct other color. The retained instrumentation measures that channel on nine occupied and 891 padded positions. All outputs are reported finite. These probes are not official-generator draws; their inputs differ from the actual retained default profiles.

\begin{table}[h]
\caption{Recorded P002 channel probes, not executions of the new witness. Zero blank positives concern the designated channel. Its count-three response fills all nine occupied positions.}
\label{tab:historicalprobes}
\centering
\begin{tabular}{@{}clrr@{}}
\toprule
Count $n$ & Other counts & Occupied minimum = maximum & Positive live / blank \\
\midrule
1 & Eight ones & $0$ & $0/0$ \\
2 & Seven ones & $0$ & $0/0$ \\
3 & Six ones & $7.862685683326549\times10^{-42}$ & $9/0$ \\
\bottomrule
\end{tabular}
\end{table}

The count-two probe has a unique mode but misses all nine required positives under an unrestricted mode-fill rule. It is a recorded input-domain limit, not a failure on Task 129's count-three contract. The summary does not recover every other channel, so it does not supply an exact full-output mismatch total for that probe. P001's earlier receipt instead records a full match on its first stored example and then 265/265; it has no corresponding amplitude table. P001/P002 identities join returned paths to retained file bytes, and P002 to the final archive; the original receipts do not themselves attest full hashes. Reviewers can inspect the protocol and returned numbers; the additional source-holder replay checks P002 on the retained corpus, not this historical probe batch or the authenticity of the joins.

Table~\ref{tab:feasibility} preserves the primary-run partition separately from the worker batches above.

\begin{table}[H]
\caption{Joint-feasibility partition for Task 053 (downward pixel shift): 105 distinct primary candidates in one nine-cycle run, with cost cap $\leq 1$. Rechecks and earlier research are excluded; the last cycle overlaps the successful worker's dispatch.}
\label{tab:feasibility}
\centering
\begin{tabular}{p{0.54\linewidth}rr}
\toprule
Outcome cohort & Candidates & Stored-case result \\
\midrule
Runtime/session rejection & 53 & Not established \\
Interface rejection & 1 & Not established \\
Runtime-valid, correct, over cost cap & 12 & 60/60 each \\
Runtime-valid, cost-eligible, incorrect & 39 & 0/60 each \\
Joint target satisfied & 0 & --- \\
\midrule
Total distinct primary candidates & 105 & \\
\bottomrule
\end{tabular}
\end{table}

The successful worker batch returns 7 min 18.484 s after dispatch; substantial earlier work makes this a closing interval, not discovery cost.

For F1, the nine late cycles contain 118 proposal rows, 106 proposed artifact hashes, and 105 distinct primary candidates; another 27 records are rechecks. Of the primary records, 84 originate in a built-in generator, 17 in replay, and four in an inbox. These provenance channels are not LLM-call counts or experimental search-strategy arms; earlier work is outside the bounded denominator.

Cycles 1--8 returned before worker dispatch; cycle 9 launched and returned during that worker's investigation. The pre-dispatch 92 primary records partition into 46 session failures, one interface failure, 12 correct over-cost candidates, and 33 cost-eligible wrong candidates. The worker's three probe receipts occur at 19:50:10.403, 19:50:56.766, and 19:52:26.191 UTC on July 17. Their order is directly recorded; the lesson about a censored semantic question is our interpretation. The later retrospective's diagnosis of conventional-traversal restrictions is weaker evidence of what every prior worker believed.

The ingested F1 pivot binds its claim to the specified software stack, complete thresholded output, and inactive-region encoding, and records rejection under another runtime version. Its falsifier names same-environment failure of the stored correctness/cost claim, an incompatible strict output interface, or a legal scorer input violating the claimed spatial map. Its scope is 60 stored cases and its status provisional. These conditions can be inspected as an empirical claim without providing a runtime construction recipe. We do not have a complete graph-retrieval--decision--gain chain for the later adaptations.

\section{Secondary screen denominators}
\label{app:screenbounds}

The finite eight-task F1 screen considers 33,867 semantic routes, yielding 270,936 first-example relation--task challenges. Only the known seed has full-stored survivors. These are semantic checks, not that many program executions. Two fixed artifacts are separately screened against 400 recipients each: the crop artifact has two first-example hits and only its own full-stored hit; the rotation artifact has three first-example hits and full-stored hits on its two already-known recipients. Thus 795 screens reject early, two reject later, and three pass the stored corpus. No eligible-mechanism recipient denominator or independent transfer success rate follows from this coverage audit.

The final post-reset lane state contains 98 passing validations on known Tasks 053, 087, and 140, 120 checker/session rejections, 44 mismatches, and two crashes. It adds no task. An earlier smoke test and pilot precede the reset, and work overlaps the original recipient adaptations; the 264 validations are neither lifetime lane effort nor a clean subsequent broad-search treatment.

\section{Present-day fixed-artifact validation}
\label{app:replay}

After resolving the campaign's reorganized paths, we selected the exact final 400-program ZIP and replayed only its nine archived perfect-task members on September 8. Archive, member, task, and retained-helper hashes were fixed before execution and checked afterward. No program was rebuilt, searched for, repaired, banked, or submitted. Each task ran in a fresh, non-root ARM64 container with network disabled, read-only source mounts/root filesystem, dropped capabilities, one CPU, 2 GiB memory, a PID limit, and a 60-second timeout. All nine processes exited successfully; none timed out.

The existing container supplied OS libraries; read-only mounts supplied Python 3.13.9, NumPy 2.4.4, ONNX 1.21.0, ONNX Runtime 1.24.4, and protobuf 7.35.1. ONNX Runtime used its CPU provider, one intra/inter-op thread, and disabled graph optimization. These package labels match the project environment observed near F1's July execution, not an immutable reconstruction of the entire historical machine. F2 also passed in this replay despite the different ONNX Runtime version linked to its historical episode.

To avoid notebook dependencies and side effects, an adapter compiled six whitelisted function ASTs unchanged from the retained official helper: sanitization, input encoding, execution/strict-positive decoding, memory accounting, parameter counting, and scoring. The original helper hash and function-AST fingerprints were recorded. The adapter applied whole-array exact equality to every retained train/test/ARC-GEN case, profiled the sanitized model for cost, and rejected skipped/error cases as evidence of a complete pass. It did not use the banking wrapper, optional zero probe, new generated cases, or an exploratory model construction. The original source files remained unchanged.

Table~\ref{tab:replay} reports the result. This is fresh local execution evidence, not a re-estimate of historical task-discovery times or recovery of private platform tests. It does not validate every other archive member, every legal input, or arbitrary software stacks. The original nine-program runner used mounted host dependencies; a separate source-holder replay below removes that installation dependency for P002 only.

\begin{table}[h]
\caption{New isolated replay of fixed final-archive members. All outputs match the full retained encoded targets; every row has zero skipped cases, wrong cases, or evaluation exceptions and receives 25 points.}
\label{tab:replay}
\centering
\begin{tabular}{lrr}
\toprule
Task and description & Passed / evaluated & Accounted cost \\
\midrule
067: repeated-panel extraction & 266/266 & 0 \\
179: grid transpose & 267/267 & 0 \\
241: diagonal reflection & 266/266 & 0 \\
053: downward pixel shift & 60/60 & 0 \\
087: 180-degree grid rotation & 266/266 & 0 \\
129: most-frequent-color fill & 265/265 & 1 \\
135: upper-right crop & 266/266 & 0 \\
140: sparse 180-degree rotation & 265/265 & 0 \\
326: upper-left crop & 266/266 & 0 \\
\midrule
Total & 2187/2187 & --- \\
\bottomrule
\end{tabular}
\end{table}

\paragraph{Additional source-holder replay of the actual P002.}
The supplement's \path{replay/} directory provides the original graph's complete equation, operand list, initializer bytes and hash, plus a fixed-model runner and dependency recipe. The 681-byte graph has one \texttt{Einsum}, 24 input references and 26 references to one length-one float32 initializer. Its count factors use distinct summed spatial indices; the remaining input supplies occupancy. Reusing the scalar adds no extra stored parameter. This describes the actual construction rather than regenerating a substitute from its idealized formula.

A newly prepared ARM64 image, with dependencies installed inside it, passed all 265 retained cases at memory zero and one parameter. It pins Python 3.13.9, ONNX 1.21.0, ONNX Runtime 1.24.4, NumPy 2.4.4, protobuf 7.35.1 and five additional dependencies. The runner accepts only the three hash-matched model/task/helper files, uses unchanged selected helper functions, and checks full decoded arrays and cost. It enforces CPU-only execution, disabled graph optimization, one thread, no network, non-root execution, read-only input mounts and resource limits. The receipt records the image, versions, architecture and cleanup. No host Python installation is mounted; no discovery, builder, tuning or new-case generation runs. This is not a cross-architecture result or an immutable reconstruction of July's environment.

\paragraph{Credential-free reviewer acquisition and tested execution.}
The additional \path{replay/reviewer_inputs.py} command reconstructs the exact 681-byte P002 from disclosed metadata using a fixed serializer. It fetches a revision-pinned public archive and helper, selects only the exact Task 129 member, and checks source sizes and hashes. Three documented helper differences affect only its module docstring and final whitespace; restoration preserves its executable syntax tree and recovers the retained whole-file hash. The routine performs no search, tuning, native execution or automatic fallback. Its default mode prints a plan; fetching requires \texttt{-{}-fetch}. Public URLs and acquisition commands are in the supplement's replay guide.

On September 9, a fresh extraction of the supplement prepared all three exact inputs without author credentials or the campaign tree. The unchanged runner then passed 265/265 complete outputs at cost one in the prepared ARM64 image, with no errors, skips or modified inputs; cleanup succeeded. The supplement's \path{replay/REVIEWER_VALIDATION.json} records both steps, code hashes and environment. This is an author-run test of the reviewer route, not an independent replication. Original inputs remain unbundled; model reconstruction and pinned public retrieval remove the private-file dependency, not upstream availability, runtime sensitivity or applicable source terms. Dependency preparation and acquisition require downloads; evaluation runs without network access. Neither replay identifies which search change produced the construction or whether a policy would rediscover it.

\paragraph{Paired native precision diagnostic.}
\label{app:precision}
On September 9 at 15:03 UTC, we ran one prespecified comparison in the same prepared ARM64 image. Unlike the helper-sanitized scoring replay, the baseline session consumed the unchanged 681 P002 bytes directly. The diagnostic derivative changes only the input/output/scalar types and exactly promotes the stored scalar to binary64; the returned binary64 array is converted once to binary32 outside the graph. The equation and all 50 ordered operands remain fixed. No coefficient, order, case or environment was searched. Both sessions use the CPU provider, one thread and disabled graph optimization; the container is non-root, network-isolated and resource-bounded. No new cost or eligibility claim is made for either this execution or the derivative.

For the 195 cases with profile $(3,2,2,1,1)$, the derivative has 18 extra positive entries each; for the 70 with profile $(3,2,1,1,1,1)$ it has nine each, totaling 4,140. P002 passes all 265 complete targets and the derivative passes none; neither loses target positives or adds positive padding. Raw binary32 outputs agree at every stored-case entry with the respective declared predictions: $(0,0,5611)u$ for P002 and $(0,1,5611)u$ after converting the derivative, for counts $(1,2,3)$. Three separately authored count probes also match both complete 9,000-entry predictions. Those probe grids are not recovered historical inputs or new independent task cases.

\path{replay/PRECISION_VALIDATION.json} records the program/script/input identities, complete-output aggregates, mapped runtime-object hashes and observed environment. The library reports version 1.24.4 and embedded commit prefix \texttt{2d924974ef}, consistent with the cited source prefix; that label is not an authenticated build. Caller-thread rounding codes and uninterpreted environment fingerprints are recorded, not a decoded kernel-state trace. The private raw arrays are retained and hashed but unbundled; the public runner can regenerate them from the exact-input route. Default verification checks the receipt and source calculation without executing either graph. This is author-run corroboration under one current stack, not independent replication. Binary64 remains finite precision and can select a different typed kernel; the result does not identify the native intermediate sequence or strengthen historical causal inference.

\section{Historical model and execution environment}
\label{app:environment}

The Task 053 closing launch explicitly selects the project virtual environment. Nearby retained probes report Python 3.13.9, ONNX 1.21.0, ONNX Runtime 1.24.4, and NumPy 2.4.4. The Task 129 official, generated-case, and bank calls instead invoke bare \texttt{python3}, and their scripts import ONNX Runtime directly. A historical interpreter comparison reports ONNX 1.19.0 and ONNX Runtime 1.27.0 in the default environment; a later probe again confirms its ONNX version. Association of F2 with that runtime is a linked inference, not an exact environment attestation inside each validation process. We do not infer F2's environment solely from the repository requirements file.

Both retained validation paths explicitly select CPU execution. Historical memory-status results report 121 GiB of OS-visible total memory, not memory exclusively consumed by the campaign. CPU model and core count, operating-system/kernel version, package wheel hashes, immutable container image, hardware expenditure, and complete local computation are not recovered. Present-day machine configuration is not substituted for historical evidence.

Selected closing turn contexts identify \texttt{gpt-5.6-sol}, with \texttt{ultra} reasoning effort for the F1 coordinator/worker and \texttt{high} for the F2 coordinator. The selected session headers record Codex client version 0.144.4. These are recorded service/client labels, not a complete all-model roster or proof of the provider's underlying implementation. The external Gemini attribution is operator-reported and is not an independently observed provider call.

\paragraph{Asset provenance and release scope.}
Original ARC is credited to Chollet~\citep{arc}; its upstream \href{https://github.com/fchollet/ARC-AGI/blob/master/LICENSE}{license is Apache-2.0}. The retained NeuroGolf helper credits Google LLC (2026), declares Apache-2.0, and has a latest version-history entry dated May 14, 2026. ARC-GEN credits Moffitt~\citep{arcgen} and \href{https://github.com/google/ARC-GEN/blob/main/LICENSE}{declares Apache-2.0}; the retained checkout is revision \texttt{a15cbdb44c77}. A retained checkout identifies an inspected asset, not the revision used by every historical process. These facts do not establish terms for all competition-distributed data or imported starting solutions, whose inventory remains incomplete. The supplement contains derived metadata, selected excerpts, analysis/witness code, a tested public-input acquisition route, fixed-program replay runner and dependency recipe. It excludes original task grids, helpers, generators, historical candidate programs, model weights and raw wellbore records. The accompanying arXiv ancillary files provide the unchanged evidence snapshot and a release notice recording the author's public-distribution authorization and CC BY 4.0 terms for author-controlled material. No upstream or separate historical-package license is assigned to it by inference.

\section{Campaign resources, progress, and portfolio integration}
\label{app:campaign}

These supporting analyses retain the campaign-level evidence; the body centers the numerical anchor and evidence-to-action episodes rather than treating campaign phases as experimental arms.

We use \emph{grinding} for repeated task-directed candidate improvement, \emph{mechanism investigation} for work intended to explain and exploit a reusable construction, and \emph{transfer} for its application to another task. These describe the work's intended object, not whether it succeeds. Grinding can include substantial rewrites, and mechanism investigation includes unsuccessful searches. Historical activities can mix these purposes; we do not force every event into mutually exclusive experimental conditions.

The operator reports \$800 in full-month payments: two Claude Max 20x and two ChatGPT Pro accounts, each \$200. This is access payment, not measured inference consumed by the roughly two-week project. External research imports attributed to Gemini 3.1 have unrecovered costs, and public assets and existing tools are part of the starting resource base. Appendix~\ref{app:cost} distinguishes payment from illustrative calendar allocation. Neither total economic cost nor a general low-budget success rate is established.

Table~\ref{tab:resources} summarizes the resource envelope and distinguishes the deadline result from later research. The competition gain belongs to the mixed campaign, including public imports and different optimizers, rather than an isolated grinding treatment.

\begin{table}[H]
\caption{Campaign resources and outcomes. Cash and approximate duration are operator-reported. Authenticated platform history corroborates both score comparisons; the competition endpoint matches the frozen leaderboard. The phases have different objectives and starting artifacts.}
\label{tab:resources}
\centering
\begin{tabular}{p{0.49\linewidth}p{0.43\linewidth}}
\toprule
Quantity & Retained or reported value \\
\midrule
Human operators / monthly access payments & 1 / \$800 for four subscription-months \\
Competition score & 7240.26 $\rightarrow$ 7655.34; $+415.08$ \\
Perfect-score tasks during grinding & 3; discovery attribution operator-reported \\
Post-deadline perfect-task additions & 6, across 2 mechanism families \\
Post-deadline artifact comparison & 7655.83 $\rightarrow$ 7672.16 \\
Gross additions / rollback loss / net & $+17.8045$ / $-1.4771$ / $+16.3274$ \\
Reported competition / later discovery interval & Approximately 213.5 h / 51 h 2 min 12 s elapsed; not human labor \\
\bottomrule
\end{tabular}
\end{table}

The later package starts from a distinct 7655.83 artifact and yields approximately 16.33 additional points after seven conservative runtime rollbacks. Retained completed platform receipts have matching public and private scores for each final package. The later result is not a competition placement, and nine is the archived perfect set, not a proof that no other task can be perfect. These endpoints establish what was achieved; the retrospective analysis next asks what the intervening records can explain.

\paragraph{Evidence and scope.}
We analyze retained candidate cycles, tool returns, validation summaries, bank promotions, manifests, and platform receipts without rerunning discovery. A separate present-day replay validates nine fixed archived programs (Appendix~\ref{app:replay}); the newly runnable scalar witness tests a different, explicitly specified proposition. Historical joins use artifact identities, timestamps, and call/return linkage. The supplement exposes selected exact excerpts, labeled paraphrases, relative chronology, and checks, but not complete private logs. Hashes establish snapshot consistency, not independent source authenticity.

The closing contexts record GPT-5.6-Sol; the two families use different launch paths and cannot be assigned one uniform runtime from a requirements file. Appendix~\ref{app:environment} distinguishes observed environment/model labels from linked inference and missing inventory.

Effort accounting covers competition-associated work through July 15 UTC, then only discovery, transfer, and relevant validation until the ninth task's bank promotion on July 18 at 03:02:12 UTC. Later packaging, retrospective writing, and generalization are excluded from time to find the set. Their retained results can still corroborate outcomes when explicitly labeled. The approximately 213.5 competition hours are a reported rounded elapsed duration whose exact absolute start is unresolved; we do not substitute the earliest surviving input as that start.

\paragraph{Units and denominators.}
We distinguish proposals, serialized candidates, evaluations, rechecks, semantic screens, fixed-artifact recipient screens, and new tasks. A passing canary is not a newly solved task; a checkpoint after reset can omit earlier work. We retain negative cases within recovered denominators and do not sum unlike units into an overall attempt count. The two family histories account for all six new perfect tasks in this archive, not a random sample of research episodes.

For competition progress, we match computed SHA-256 hashes for all 400 programs in the exact starting and deadline ZIPs to the ledger's retained 16-hexadecimal-character prefixes, without executing them. The primary comparison requires an initially accepted import and at least one matching deadline-artifact record marked both \texttt{ok} and \texttt{banked}; a promotion flag alone is insufficient. This acceptance-record cohort has 394 tasks, not a guarantee that every matching record passed. Thirteen also have \texttt{ok=false} records; we report a sensitivity excluding them. The full 400-task cost table is descriptive metadata, not 400 independently revalidated programs. Selected archives, not the evolving bank's historical maxima, define the endpoints; comparisons of unchanged archive members use full hashes.

\subsection{Three perfect tasks during grinding, alongside broad portfolio gains}

The starting archive came from franksunp's NeuroGolf Variant B Mark A notebook~\citep{initialbaseline}; its version-58 producer record reports 7240.26, and its ZIP matches all first-import artifacts. An authenticated September 8 platform audit separately confirms 7240.26 for the operator's first visible submission and 7655.34 for the frozen leaderboard entry, matching the retained competition package's timestamp. The displayed gain is 415.08. A later pre-midnight receipt scores 7655.38 but is not the frozen leaderboard record (Appendix~\ref{app:platform}). Table~\ref{tab:competition} analyzes the exact retained 7655.34 package, not the maximum of every submission receipt.

\begin{table}[H]
\caption{Starting-to-deadline artifact comparison. The 394-task cohort requires local acceptance records at both endpoints. The final column additionally excludes 13 tasks with mixed success flags for the deadline artifact. Neither cohort is a fresh validation; cost-implied gains do not reconstruct platform grading.}
\label{tab:competition}
\centering
\begin{tabular}{@{}p{0.36\linewidth}rrr@{}}
\toprule
Quantity & All tasks & \shortstack{Acceptance\\records} & \shortstack{No mixed\\records} \\
\midrule
Compared tasks & 400 & 394 & 381 \\
Lower recorded cost & 353 & 348 & 335 \\
Unchanged recorded cost & 47 & 46 & 46 \\
Higher recorded cost & 0 & 0 & 0 \\
Cost-implied point change & $+415.1582$ & $+409.2778$ & $+395.3424$ \\
\bottomrule
\end{tabular}
\end{table}

Within the acceptance-record cohort, 348 tasks improve and 46 remain unchanged. The ten largest changes contribute only 9.32\% of that cohort's cost-implied gain. Broad improvement survives exclusion of the 13 mixed-outcome tasks; four have a failed or incomplete record after their first accepted promotion, with the cause unresolved. The grinding-phase perfect tasks are 067 (repeated-panel extraction), 179 (grid transpose), and 241 (diagonal reflection).

The operator attributes all three discoveries to grinding. Local construction and successful cost-zero evaluation are corroborated for 179 and 241. Yet all three selected programs still match the public comparison archive: a strict-improvement bank rejects equally cheap replacements, so a local success need not change the incumbent. For 067, the located later ``rebuild'' is an incumbent recheck; its origination episode is unrecovered. The comparison archive was acquired after campaign activity began, not at a verified pre-campaign discovery baseline. Author-reported discovery, observed construction, and selected-artifact ancestry therefore remain distinct; unchanged bank bytes do not imply an absence of discovery.

The archive join connects these gains to the later histories. Five later perfect-task additions retain their exact starting artifacts at the deadline: 053 (downward pixel shift), 087 (180-degree grid rotation), 129 (most-frequent-color fill), 140 (sparse 180-degree rotation), and 326 (upper-left crop). Only 135 (upper-right crop) already improves, from cost 200 to 30. All six deadline costs equal their costs in the later 7655.83 starting package. Endpoint equality does not establish that a task received no work or that only discovery could improve it. The short task labels describe their grid transformations, not implementation recipes.

The all-task cost-implied delta differs from the displayed platform delta by 0.0782 points; we do not force agreement between local metadata and private platform grading. Public imports and multiple optimizers also preclude assigning the gain to grinding alone.

\paragraph{Conclusion and next question.}
In the unclipped scoring range, reducing cost from $a$ to $b$ adds $\ln(a/b)$ points: many multiplicative reductions can accumulate substantial gain. For already accepted programs, perfect-task count changes only when a task reaches cost at most one. A campaign can therefore deliver widespread value without changing that count. The observed 348 improvements support cumulative optimization as an important achievement; the unchanged five later targets motivate a different question, not a claim of zero progress: \emph{which requirements were preventing a candidate from becoming perfect?}

\subsection{Realized gains required validation and portfolio protection}

The six new tasks contribute 17.8045 gross points over the distinct post-deadline starting artifact. Seven conservative runtime rollbacks on other tasks surrender 1.4771 points, leaving 16.3274 net. These are artifact changes, not an estimated probability of runtime failure. They demonstrate that adding perfect tasks and protecting a usable portfolio are separate responsibilities. Reporting only the six additions would omit a cost paid in the realized package.

Different validation records also have different scopes. An intermediate Task 135 artifact passes only 320 of 5,000 cases from an external generator. The later, differently hashed artifact passes the retained executable checks using the competition generator, including separate isolated fresh-case runs. Because both the artifact and generator differ, this is a failed validation branch in the history, not evidence that the banked artifact failed its official gate or that either generator is generally superior.

Stored-corpus checks, freshly generated tests, and platform grading are separate evidence layers. The stored Task 053 corpus has 60 released examples; Task 129 has 265. Generated cases can test additional behavior but are not automatically platform-hidden examples, and a custom sampler need not match the official generator's distribution. A bank entry is local acceptance; the later complete-package platform receipt is separate outcome corroboration, not a per-task formal certificate.

A separate author-run replay re-executed the nine fixed archived programs on all 2,187 retained cases: every case passed and every program received 25 points, with eight costs of zero and Task 129 at one. This confirms current finite-corpus behavior without reconstructing discovery or the private platform tests (Appendix~\ref{app:replay}).

For Task 129, public material identifying the missing known-perfect target is retrieved before an operator requests that the answer not be inspected. The subsequent instruction is evidence of intended procedure, not proof of no earlier exposure. We describe the episode as targeted reconstruction following public existence information, with independence of solution details not fully audited. Its final bank promotion ends our discovery-effort window. A 36-candidate F2 generalization manifest and later package grading occur after that endpoint; neither is charged to time required to find the archived set.

\paragraph{Conclusion.}
The durable outcome is the accepted \emph{portfolio}, not the sum of celebrated candidate gains. The operational implication is to return validation outcomes and rollbacks to shared memory, so future workers can retrieve both a useful construction and the evidence limiting where it should be trusted. Complete propagation of every historical update is not established here.

\section{Subscription payment and calendar allocation}
\label{app:cost}

The operator reports two Claude Max 20x and two ChatGPT Pro subscription-month payments, each \$200: \$800 in full-month access payments. The operator describes the project as roughly two weeks; the approximately 213.5-hour competition campaign and the later 51 h 2 min 12 s discovery interval are separately reported in Appendix~\ref{app:campaign}. These descriptions do not identify all account billing dates, access outside this project, or per-provider inference usage.

For transparent comparison only, uniform time allocation of these payments over a hypothetical 30-day month would assign
\begin{equation}
 A(D)=\$800\,\frac{D}{30},\qquad A(14)=\$373.33.
 \label{eq:allocation}
\end{equation}
Equation~\ref{eq:allocation} is an accounting convention, not a measured expense reduction: subscriptions are paid in full, quota use can be concentrated, and unused days need not be refunded. It is an illustrative allocation, not an estimate of API-equivalent consumption or the marginal cost of later discovery. Unrecovered external-provider charges, hardware, electricity, and human time prevent a complete project-cost total. The \$800 figure is consequently neither an all-provider audit nor a claim that every researcher can reproduce the result at that price.

\section{Conditional human-effort reconstruction}
\label{app:effort}

The scoped corpus has 444 competition inputs (317 Claude, 127 Codex) and 39 post-deadline discovery inputs (all Codex). We count interactive operator event records once; worker messages, duplicated message representations, and later replayed forks are excluded. Five intervening non-discovery inputs remain barriers rather than disappearing from the chronology. The retained clients do not cover all off-client thinking, preparation, reading, or terminal activity.

The initial pattern classifier is rule-based. A targeted audit selects 50 inputs, including all initially long/imported items and selected ambiguous/design classes. Forty-nine request framings are reviewed; one sensitive item remains metadata-only. Thirty category assignments change. Thirteen reviewed inputs retain alternative plausible categories. The 433 unselected inputs and one metadata-only input together leave 434 of 483 inputs without a framing review. This is LLM-assisted semantic interpretation, not independently coded human ground truth or a random validation sample. The resulting primary labels comprise 150 orchestration, 92 operational, 58 design/correction, 57 monitoring, 42 commands, 11 long/imported, seven acknowledgment, and 66 ambiguous inputs.

For input $i$ and scenario $s$, a pattern-specific base duration $B_{is}$ represents preparation, deciding, and interface actions. When a new visible response exists, assumed reading duration is
\[
R_{is}=\min\{w_i f_s/r_s,\; k_s,\; a_i\},
\]
where $w_i$ is the latest visible response's word-unit count, $f_s$ the assumed read fraction, $r_s$ the reading rate, $k_s$ a cap, and $a_i$ the time since that response appeared, in consistent minute units. Missing Claude responses use an explicit imputed reading duration; commands have no added reading term. Thirteen Codex inputs lack a new visible response and receive no new-response reading term, without assuming no other reading occurred.

We place $B_{is}+R_{is}$ in a window ending at the input, clip it at phase and explicit exclusion barriers, and sum the union of windows across clients. This avoids charging overlapping assumed activity twice and gives isolated inputs nonzero modeled effort. Window placement, durations, and reading behavior are assumptions, not observations of attention. Unknown preparation of the two Gemini imports is not reconstructed from pasted length.

\begin{table}[h]
\caption{Analyst-chosen base minutes per input. These settings are neither calibrated rates nor statistical priors learned from the records.}
\centering
\begin{tabular}{lrrr}
\toprule
Pattern & Low & Middle & High \\
\midrule
Command & 0.10 & 0.30 & 1 \\
Monitoring & 0.25 & 1 & 3 \\
Acknowledgment & 0.15 & 0.50 & 2 \\
Operational & 0.50 & 2 & 8 \\
Orchestration & 0.75 & 3 & 10 \\
Design/correction & 2 & 8 & 25 \\
Long/imported & 0.50 & 2 & 6 \\
Ambiguous & 0.50 & 2 & 8 \\
\bottomrule
\end{tabular}
\end{table}

Low/middle/high reading settings use 300/200/120 word units per minute, read fractions 0.25/0.5/1, caps 2/6/15 minutes, and missing-response allowances 0.25/1/3 minutes per eligible input. Together they yield 6.77/24.07/59.97 modeled hours. Middle-scenario components are 22.27 competition and 1.81 discovery hours after independent rounding; these need not sum exactly to the rounded total. Varying only missing-response reading to 0/1/3/5 minutes changes the middle total to 20.31/24.07/30.75/36.28 hours. Changing the 13 declared alternative categories together yields 24.12 hours. These sensitivity checks do not validate the remaining labels or the duration assumptions.

An alternative descriptive calculation sums consecutive pooled inter-input gaps only when they are at most 5/15/30 minutes. Its totals are 7.77/27.35/50.66 hours, excluding singleton padding. It measures a different proxy; proximity to a pattern-model result is not corroboration. Neither analysis supports dividing phase outcomes by inferred labor to establish relative research efficiency.

\section{Conditional knowledge and task descriptions}
\label{app:knowledge}

The retained pre-breakthrough base knowledge graph has 745 nodes and 4,071 edges; a source-preserving overlay has 880 nodes and 5,691 edges. Model/operator records cover all 400 tasks. Semantic annotations are narrower: 36 feature nodes and 59 task-feature edges cover 19 selected tasks, alongside nine motifs with preconditions and ten manually specified transfer candidates. Relation types distinguish task features, constraints, implementation operators, semantic analogies, tested successes, failures, incompatibilities, and proposed transfers. Operator overlap alone is therefore not asserted to establish semantic transfer.

Candidate ranking uses estimated logarithmic score gain multiplied by manually assigned confidence. This documents a structured transfer worklist, not calibrated probabilities or a generic graph-inference method. The associated reports find no new perfect task. Retained event timestamps place both graphs before the post-deadline family discoveries, but do not establish when their concepts were first devised or a causal effect on later work.

The Task 053 loop separately requires knowledge records to declare a kind, scope, family, mechanism, statement, assumptions, falsifier, and supporting artifact. It checks artifact identities and prevents an existing claim identifier from silently changing meaning. The post-success integration cycle actually ingests one provisional pivot with four assumptions and bounded stored-test scope. A graph edge or claim label denoting a success is not a formal proof; its support remains the specified validation record.

The short task labels in the paper are descriptive names checked against the official generators, not official benchmark titles. Task 053 shifts pixels down one row in a $3\times3$ grid; 087 rotates a $3\times3$ grid by 180 degrees; 129 fills a $3\times3$ grid with its uniquely most frequent color, which need not be a majority; 135 extracts the upper-right $3\times3$ block of a $9\times9$ grid; 140 rotates a sparsely populated $3\times3$ grid by 180 degrees; and 326 extracts the upper-left $2\times2$ block. The grinding-phase tasks are 067, extraction of a repeated square panel; 179, grid transposition; and 241, reflection across a marked diagonal. Task 385 completes a vertical mirror pattern. These descriptions identify the semantic problems without providing optimized construction recipes.

\paragraph{What the knowledge graph retained.}
One retained rule is concrete: a static crop requires a fixed source region and orientation, with no input-dependent object selection. A recipient that must first identify an object needs a separately validated selector; sharing the final crop operation is insufficient. This illustrates recorded knowledge, not a rule credited with selecting the later F1 adaptations. We recover pre-success report access for F2 and post-success pivot ingestion for F1, but not a graph query--retrieval--adaptation--gain chain. Memory consolidation and memory-enabled discovery remain distinct claims.

\paragraph{What transferred.}
Family F1 contributes Task 053 and four recipients. Its common object is a task-adapted mapping or aggregation of spatial support: shifts, rotations, and crops can share an idea without implementing the same input--output map. F2 instead discriminates global color multiplicity using finite-precision behavior for Task 129. The families follow construction histories and adaptation records, not equal scores. Tasks 087 and 140 share a successful artifact identity; this limited sharing does not imply universal reuse. Table~\ref{tab:transfer} summarizes the retained family outcomes.

\begin{table}[H]
\caption{Family outcomes and selected adaptation records. Stored passes refer to banked artifacts; the initially failed and later successful adaptations are not independent experimental conditions. Secondary screening denominators are in Appendix~\ref{app:screenbounds}.}
\label{tab:transfer}
\centering
\begin{tabular}{p{0.40\linewidth}p{0.52\linewidth}}
\toprule
Outcome or recovered stage & Result and scope \\
\midrule
053: downward pixel shift & F1 seed; cost 0; 60/60 stored cases \\
135: upper-right crop & F1 adaptation; cost 0; 266/266 \\
087: 180-degree grid rotation & F1 adaptation; cost 0; 266/266 \\
140: sparse 180-degree rotation & F1 adaptation; cost 0; 265/265 \\
326: upper-left crop & F1 adaptation; cost 0; 266/266 \\
129: most-frequent-color fill & F2 seed; cost 1; 265/265 \\
\bottomrule
\end{tabular}
\end{table}

Task 087 makes adaptation visible: its first selected rotation attempt passes 0/266 stored cases; a later construction passes 266/266. Static reconstruction of the recorded spatial maps finds the intended rotation in both, but repeated positive support outside the logical $3\times3$ region only in the failed form. This is a full-output distinction, not merely recognition of the rotation rule, and not an isolated runtime ablation. Task 140 follows the same path with 265 cases. The supplement links later hashed validation, promotion, and shared final bytes, while leaving the early in-memory candidates' byte identities unknown. Task 385's selected mirror-completion attempt passes 0/265; this bounds a failed branch, not the task's applicability.

Broader fixed-artifact screens and a post-reset validation lane add no new task. They measure coverage and repeated validation, not the success rate of task-adapted transfer; their work overlaps the original adaptation path. Appendix~\ref{app:screenbounds} retains all denominators and negative outcomes. There is no clean targeted-versus-broad experiment.

\section{Independent platform-history corroboration}
\label{app:platform}

On September 8, we queried the authenticated competition-submission API read-only through all three returned pages: 422 unique records, with current statuses of 398 COMPLETE and 24 ERROR. The first visible operator submission is dated July 7 at 02:25:50.273 UTC and scores 7240.26. This is distinct from the earlier public producer's submission, although score and reported file size agree with the retained seed. The list API provides no server-side content hash; local hashes and retained manifests remain the source of archive identity.

The live frozen leaderboard returns 7655.34 and the July 15 23:56:33.053 submission timestamp, matching the analyzed competition package. Another receipt dated 23:59:53.450 now shows COMPLETE and 7655.38, but is not the frozen leaderboard entry. The API does not explain that distinction or supply historical grading-completion times. We therefore retain the frozen entry rather than replace the analyzed archive with the highest pre-midnight receipt. Exactly 395 records currently marked COMPLETE have submission dates through the frozen entry; 396 precede midnight when the later receipt is included. Neither count establishes how many gradings had finished by the deadline.

The same live history corroborates the distinct post-competition baseline at 7655.83, submitted July 16 at 00:33:03.297, and the 7672.16 package, submitted July 18 at 04:04:33.597. Public and private scores agree for all five named receipts. SDK timestamps are interpreted as UTC from their exact agreement with retained UTC-marked receipts; submission dates are not completion dates. The full private audit records pagination, identities, sizes, and hashes, while the anonymous supplement omits account-linked receipt identifiers. This new platform check strengthens endpoint evidence, not per-task correctness, original-discovery attribution, or the causal contribution of a search mode.

\section{A retained cross-domain comparison: wellbore prediction}
\label{app:crossdomain}

We additionally inspect one retained ROGII wellbore-geology replication episode from the same investigator. This is a selected historical comparison, not a new training experiment, an independent operator replication, or part of NeuroGolf's effort interval. The companion \path{cross_domain/} packet links 48 selected fields to 15 source hashes. Its verifier checks retained arithmetic and chronology; author-side source checks re-extract those fields without importing model code. The packet does not recompute scores from raw predictions or authenticate the complete training history.

A prior check found agreement within $10^{-6}$ on 16 mapped evaluation-input channels over 25 wells. A subsequent brief generalized this to complete input exactness. Two distinctions limited that inference. First, the check concerned selected version-1 channels, not the whole 20-channel carrier or a later version-2 feature. Second, evaluation-time agreement did not establish agreement under training augmentation. A proposed explanation---moving the target while keeping observed gamma ray---was itself rejected: the old implementation already regenerated gamma ray.

The surviving diagnosis concerned the augmented candidate and target-path distributions. Candidate-axis support is the fraction of candidate offsets within the reference well's coverage. Alignment margin is median candidate cost minus cost at the nearest true offset; positive values favor the true offset. In a retained 200-draw diagnostic, support was approximately 0.697 versus 0.987, with median margin $-0.0084$ versus $+0.5050$. The arms shared selected wells and integer seeds but not RNG streams, so these are distribution summaries, not equal augmented tensors. A separate acceptance batch for an augmentation-family replacement increased support from approximately 0.711 to 0.989; no-hit margin summaries remained equal, not independently verified tensor bytes.

The matched decision used full-fold CPU evaluations: control RMSE 7.1043 ft and intervention 6.2904 ft, both on 155 wells, seed/fold zero, after 30 epochs of a 300-epoch schedule. The 0.8139-ft gain failed the brief-specified strict $>1$-ft promotion rule; the report declined a 300-epoch handoff. Four-decimal display rounding cannot change this decision. The intervention changed an augmentation family and calibration, not one isolated operation. Neither an approximate older reference, a 60-well subset nor a GPU score is substituted for the matched CPU comparison.

Embedded records place the augmented diagnostic at 19:46 UTC and intervention acceptance at 20:05 on August 8. Training logs supply 20:34/20:35 clock times, with date/timezone supplied by the report; the version-scope audit is later, at 21:33. That later correction cannot explain selection of the earlier A/B. The brief's gate lacks an independently recovered creation timestamp, so it is not claimed as immutable preregistration. Ten recovered historical hash quotations match current bytes; five other sources lack such quotations.

The useful recurrence is scoped diagnosis with a consequential rejection decision. It does not establish that the NeuroGolf workflow caused the gain, that diagnosing version parity was the intervention, or that this practice improves average research efficiency. The top-team accounts cited in the body similarly supply reported convergence, not a controlled denominator or evidence of independent invention.

\input{handoff_appendix}

\end{document}

%% file: body.tex
\begin{abstract}
Autonomous coding agents can remember an experiment yet carry forward a conclusion it does not justify. We reconstruct how evidence is reused in a 400-task NeuroGolf campaign, with selected wellbore-prediction records from the same operator as cross-domain comparisons. A numerical counterexample exposes an overbroad exclusion; other episodes distinguish what failed, incomplete and revised programs justify doing next. The wellbore records reveal an additional weakness: a coordinator correctly acknowledges a novelty-only rejection, then summarizes it as measured closure. Separately, a prefix-based acceptance gate at three thresholds yields worse target scores than ungated adaptation in the retained experiment. These cases motivate an inspectable handoff linking the tested proposition, its scope, candidate and evaluator identity, check statuses, and reopening condition. A dispatch tree separates investigation, repair, reopening and stopping; a promotion predicate requires that each check both passes and has evidence supporting its use for the requested decision. The practical lesson is to preserve not just experimental results, but their limits on subsequent action.
\end{abstract}

\section{Introduction}

After an unsuccessful coding attempt, what should the next agent do? Retrying can waste effort when an applicable counterexample is already known. Stopping can waste an opportunity when the previous search tested only one representation. Both errors can occur in a system with persistent memory: the problem is not simply whether evidence is stored, but what conclusion it supports and where that conclusion is applied.

Our thesis is that \emph{autonomous research must not merely accumulate experiments; it must preserve what each experiment justifies and use that evidence to choose the next action}. A failed program, an unsuccessful family search, a component-level lower bound and an execution rejection justify different responses. One exact failing input can refute the tested program under its evaluation conditions. A finite search without a survivor need not rule out other representations. Confusing these statements can make the same system too conservative about exploration and too permissive about promotion.

We study this tension in NeuroGolf~\citep{neurogolf}, a 400-task colored-grid program-optimization competition. A \emph{perfect-score solution} earns 25/25 for correct tested outputs at accounted memory-plus-parameter cost at most one; it does not mean zero physical computation. The operator attributes three perfect-task discoveries to grinding; later work adds six across two construction families. These are the archived nine, not nine independent mechanisms or all tasks that could be perfect. Broad optimization also matters: 348 of 394 tasks with endpoint acceptance records improved during the mixed campaign (Appendix~\ref{app:campaign}). Perfect-task count alone would miss that progress.

The human role was process design and steering---improving the grinding loop, introducing mechanism-discovery work, supplying information and changing allocation---not individual solution review. Agents performed the diagnoses analyzed here, sometimes inside ordinary rebuild workers. We therefore do not compare ``human audit'' against ``agent grinding,'' or assume diagnosis begins only when a separate research mode is named.

We contribute three connected objects: (i) an executable, task-specific counterexample to an overbroad numerical rationale, with a conditional explanation of the archived program; (ii) source-linked decisions and a wider cost-state inventory that expose both useful restrictions and unsafe handoffs; and (iii) a concrete evidence-to-action protocol derived from these observations. The numerical and validation phenomena are familiar. What the reconstruction adds is their connection to the particular inference, subsequent action and validation boundary that a future research loop must preserve. The protocol is proposed practice, not a newly evaluated search algorithm.

Interpreting that evidence requires both the evaluator's definition of success and a record of how results were carried forward.

\section{Setting and evidence}
\label{sec:setting}
\label{sec:methods}

\paragraph{What the evaluator checks.}
Each task requires an ONNX program for an ARC-derived grid transformation~\citep{arc,arcgen}. The retained helper encodes grids as padded, ten-color, one-hot float32 tensors of shape $(1,10,30,30)$. It thresholds every output entry strictly above zero and compares the complete encoded target: neither argmax nor visible-grid equality suffices. Interface and execution checks also apply. For an eligible correct program,
\begin{equation}
s(c,t)=\max\{1,\; 25-\ln(\max\{1,M(c,t)+P(c)\})\},
\label{eq:score}
\end{equation}
where $M$ is accounted intermediate memory in bytes and $P$ stored parameter count in elements; incorrect or invalid programs score zero. A finite-corpus maximum is not general correctness, a runtime guarantee or a formal proof.

\paragraph{Archive-wide inventory, selected explanations.}
We read 29,591 competition ledger rows for all 400 tasks, coverage notes for 318 tasks, task/approach records, six retained cost-floor registry versions and five allocation-policy source versions. A registry ``floor'' records a cost and unsuccessful-search strikes; it is not assumed to be a proved bound. The fixed extraction rule selects the first Git-observed task/cost state with at least two strikes: 97 states across 92 tasks. The raw 128 current registry keys represent 114 tasks because 14 have padded/unpadded aliases. The packet preserves rather than silently merges these identities (Appendix~\ref{app:decisions}).

Table~\ref{tab:registry} joins every selected state to its frozen deadline artifact. It shows why a cost/strike label is not a complete decision record: lower acceptances can precede it, follow it, or fail to survive. In 29 states a subsequent lower acceptance does not leave a lower deadline endpoint; six states have later negative rechecks of a lower accepted hash. These are not 29 unsafe algorithms or 51 refuted impossibility claims. Acceptance means recorded \texttt{ok=true, banked=true}, not fresh validation, and a restriction may concern another representation. Even the 91 canonical-key-eligible states are not observed exclusions: historical scratch-registry bytes and dispatch exposure remain unknown.

\begin{table}[h]
\caption{Historical cost-state contrasts, not audit trials. The first two rows overlap; the three outcomes in the final row partition all 97 states. Deadline costs belong to frozen artifact identities, not best temporary candidates.}
\label{tab:registry}
\centering
\begin{tabular}{@{}lr@{}}
\toprule
Observation & States \\
\midrule
Lower accepted cost already recorded before observation &16\\
Lower accepted cost recorded afterward &79\\
Deadline cost below / equal to / above registered cost &51 / 35 / 11\\
\bottomrule
\end{tabular}
\end{table}

\paragraph{What chronology can establish.}
We reconstruct four additional episodes with pre-outcome reasoning and subsequent actions, alongside the previously selected perfect-score families. They are development examples; Task 368 is an outcome-informed no-improvement contrast, not a holdout. Operator requests come from interactive-root records, not automatically supplied worker \texttt{user} messages. Selected excerpts, candidate versions and receipt links are available in the PDF and \path{decisions/} packet. LLMs assisted extraction, analysis, drafting and critical review; deterministic checks establish consistency with retained sources, not independent human coding or historical authenticity. Time order does not identify causation. Mathematical diagnostics and September fixed-program replays are separate evidence types; none reconstructs discovery. Competition accounting ends after July 15 UTC, and later discovery at July 18, 03:02:12 UTC; subsequent writing and validation are not charged to discovery effort.

We also inspect selected ROGII wellbore-prediction episodes from the same operator, recorded in August before this protocol was formalized. Their linked excerpts and aggregate measurements test the scope of our interpretation, not its effectiveness in another population (Appendix~\ref{app:crossdomain}); they are outside NeuroGolf's effort interval.

Cost-state histories cannot settle whether a particular exclusion is justified. The next case compares one rationale's assumptions directly with the evaluator's behavior.

\input{numerical_anchor}

\section{What prior evidence permits the next worker to do}
\label{sec:reuse}

The numerical case requires weakening an overbroad conclusion. That is only half the problem: other episodes require preserving and applying a restriction more carefully. The spatial family supplies a parallel observation: two rotation constructions agree visibly but differ on padding that the full-output evaluator checks (Appendix~\ref{app:spatial}). Table~\ref{tab:decisions} maps five further situations to different next actions. The organization is ours; the actions and unsuccessful branches are historical, agent-led records rather than trials of that organization.

\begin{table}[h]
\caption{Different evidence warrants different action. D196--D368 link selected excerpts, program identities and receipts in Appendix~\ref{app:decisions} and the packet. No row estimates time saved or a policy's success probability.}
\label{tab:decisions}
\centering
\begin{tabular}{@{}p{0.27\linewidth}p{0.34\linewidth}p{0.32\linewidth}@{}}
\toprule
Retained evidence & Author-derived next question & Observed outcome / limit\\
\midrule
196: propagation-only rationale & Does bounded perimeter matching escape those assumptions? &2360$\to$2300; specific memory exposure unknown\\
188: factored certificate adds disagreements & Can a repaired certificate recover reference outputs? &cost-115 matches cost-129 on the paired batch; two numerical features change\\
188: worker's scoped minimality claim & Can the controller improve while the writer stays? &Different 109-cost program succeeds; equal costs are not equal bytes\\
368: rank/operator/cost constraints & Is a distinct, admissible route affordable? &Specific routes stop; 852 retained, not proved optimal\\
364: incomplete, slow or non-retained candidate & Which gate still blocks usable progress? &4668 is partial; later local 4888 not retained over 5067\\
\bottomrule
\end{tabular}
\end{table}

\subsection{Task 188: preserve a failure, localize a bound}

Task 188 recovers the original tile from a horizontal or vertical duplication. Some 4-by-4 inputs admit both explanations, so disagreement with a latent generator label is not interchangeable with disagreement against a specified reference. A worker rejects a factored cost 127 candidate after 10/2000 fresh failures despite stored-case agreement. A later worker banks the same named program, then observes 5 failures in another 2000-case batch. The first link is path/build-version evidence, not authenticated identical bytes; the second bank/fresh-check link includes the accepted hash. The record does not show that the later worker read the earlier warning before promotion.

Before its successful result, the later worker specifies a nonnegative self-product certificate and an operand arrangement intended to control cancellation. A paired batch then compares the unfactored 129, factored 127 and repaired 115 programs: they match 1999/1992/1999 latent labels and differ from the 129 reference on 0/7/0 cases, respectively. These are three different fresh batches, not inconsistent estimates from one batch. The measured repair changes both certificate structure and operand order; it does not isolate one numerical cause. Its usefulness is concrete: a targeted comparison recovers the desired reference behavior while reducing cost, rather than treating another stored-corpus pass as sufficient evidence.

Later, another worker explicitly reads the coverage note (D188-E10), then states its intended redirection before the new candidate's result:

\begin{samepage}
\textbf{D188-E08, agent assessment:} ``The rank-reduced writer is already minimal in its spatial part: two 30-cell affine factors (rank two), and sparse factor inputs cannot pass ONNX shape inference. That disproves further low-rank compression here. I found a genuine controller rewrite instead: scale the duplicate certificate so the already-computed height moment becomes its upper threshold, deleting the lone \texttt{1e8} parameter.''
\par
\end{samepage}

The new program is accepted at 109. An earlier, different 109-cost program failed 94/266 stored cases. We do not independently prove the quoted minimality claim. The observed use is narrower: the worker treats one component as fixed and redirects its proposal to the controller. The failed program's evidence remains tied to its own identity and evaluation conditions.

The system implication is not merely ``remember failures.'' At dispatch, the next worker needs the scope of the prior result; at promotion, applicable counterexamples must be available to the validator. The archive establishes a handoff risk, not that an executable regression corpus was already enforced or that enforcing one would have saved a measured budget.

\subsection{Reopening and stopping are both constructive outcomes}

Program-specific failures leave two further questions: when may another construction reopen the search, and when does a scoped constraint justify stopping?

\paragraph{196: intact-frame recoloring.}
A retained note says bounded propagation is the only legal strategy, but also permits reopening for a genuinely new construction. A later worker reads prior approaches and, before testing, proposes nine perimeter signatures for the generator's bounded 3--5 cell frame dimensions. It avoids propagation, improves 2360 to 2300, and passes 266 stored plus 2000 fresh cases. The universal architectural sentence is too broad; the conditional reopening rule is sensible. We do not observe the worker reading that specific sentence, nor infer that challenging it caused the improvement. An ambitious dispatch target also remains unjustified: a good aggregate leaderboard average does not prove an exceptionally cheap program exists for each task.

\paragraph{368: prototype copying.}
An alternative representation is not always admissible. A worker reads and applies a recorded identity-minor rank argument to a static direct-contraction construction, not to every nonlinear program. It explores alternatives, then stops routes needing excluded operators or an over-budget layout fallback and retains 852. The unchanged cost does not prove optimality or resource savings. The scoped argument and admissibility checks explain the stop. This branch matters: a diagnostic instruction that always demands another implementation can itself waste research effort.

\paragraph{364: component-shape recoloring.}
Other routes remain open in principle but blocked at a required check. A worker-reported exact NumPy classifier does not yet yield a complete, runtime-safe ONNX improvement. The nominal 4668-cost prototype leaves the U class unimplemented and its bounded run times out. Useful algebra can be retained without promoting an unfinished program; bounded repair may still be justified. A later, separate 4888 candidate is locally accepted, but the deadline artifact retains 5067. Records attribute rollback to a failed singleton submission, with conflicting control-score metadata preserved. We establish non-retention, not the exact remote rejection mechanism. These cases distinguish a failed construction, a repairable implementation and a deployment decision.

\section{Designing the loop around the next decision}
\label{sec:broad}
\label{sec:validation}

These task-level distinctions also depend on the surrounding process: what workers are asked to pursue, which contract governs success, and what the coordinator carries forward.

\paragraph{Process steering, not individual proof review.}
Two source-linked operator chains show where the research process itself can diverge from its target. A request to prevent task starvation is initially implemented as a three-day rotation. The operator corrects this to coverage before the imminent deadline; source changes implement an all-task round, including previously filtered tasks, with one decisive hypothesis and a stop. The source specifies a prompted budget of 20 minutes/about 20 calls and a separate 25-minute outer timeout, not an empirically optimal budget or verified duration for every worker. The chain establishes request-to-implementation linkage, not completed coverage or attributable gains.

Later, a discovery protocol requires agreement on all 70 generated configurations before testing 60 stored Task 053 cases. After a request to identify bottlenecks, the agent diagnoses the mismatch with the fixed-corpus objective; the revised process separates primary acceptance from broader robustness. This is not evidence that the eventual winner had already been generated and wrongly rejected. It is also not permission to weaken required deployment checks after seeing a candidate: a fixed-corpus score and general correctness are different claims. The objective, applicable contract and robustness annotation must be explicit (PCOVERAGE/PCONTRACT, Appendix~\ref{app:decisions}).

\input{handoff}

\section{Related experience and limits}

The handoff specifies what feedback-and-memory records must preserve for a subsequent decision.

Reflexion already uses linguistic feedback and episodic memory to influence subsequent trials~\citep{reflexion}; research--evolution hybrids and budget-aware agent evaluation likewise precede this study~\citep{deepevolve,emosta,agentsmatter}. We do not claim persistent feedback or combining search with research as new. Our contribution is the inspectable connection between a particular recorded restriction, its valid scope and the subsequent action or gate.

Primary NeuroGolf accounts already document numerical sensitivity, selector constructions and conditional memory~\citep{neurogolf4,neurogolf9,representation}. First-place process design reports better token use from retargeting and shared recipes~\citep{neurogolf1pipeline}; the eighth-place report describes repeated old work after blank restarts~\citep{neurogolf8}, whereas the ninth-place report records useful unattended optimization~\citep{neurogolf9}. These are practitioner accounts, not matched-budget comparisons of our proposed protocol. Other teams' direct human solution review is a different intervention from this operator's process steering.

One operator, public imports, changed information, overlapping workers and selected histories limit causal interpretation. Registry snapshots do not establish actual exclusions; a later cheaper program does not refute every earlier scoped restriction. The nine-program replay covers 2187 retained cases, not private platform inputs or arbitrary runtimes. The numerical source model is conditional, not July's execution trace. No automatic triage accuracy, complete episode-level resource comparison or general search-efficiency benefit is identified. The supplement provides safe derived records and source excerpts, not full private sessions; de-identification cannot guarantee against linkage. Author responsibility and release decisions remain separate from deterministic checks.

\section{Conclusion}

Useful autonomous research must preserve both the freedom to test an unruled-out construction and the obligation to respect an applicable failure. The numerical witness, decision episodes and process corrections show where those obligations differ. The resulting practical target is not more experiments or more skepticism alone, but evidence that changes the next proposal, validation or stopping decision without claiming more than the experiment established.

%% file: numerical_anchor.tex
\section{When a negative conclusion exceeds its evidence}
\label{sec:findings}
\label{sec:joint}

Task 129, mode-color filling, requires filling a $3\times3$ grid with its uniquely most frequent color. In the retained contract that color occurs three times and every other color at most twice. A cost-80 program already solved the stored cases; the unmet target was cost at most one. An earlier bounded search explored unweighted repeated-input relations. Its implementation already used float32 signatures, but its returned report separately dismissed a scalar extension:

\textbf{Prior returned report, X001:} ``A one-scalar initializer does not add a new threshold relation here: it can only scale or invert the monomial, so the genuine relational support delta remains cost 0.''

The flaw is the scope of that reason, not the fact that the restricted search failed. Exact positive support ignores changes in positive magnitude; the actual evaluator thresholds the result of finite-precision execution. Those are different equivalences. The later cost-one program P002 has idealized expression $o_{hw}n_c^{23}a^{26}$, where $n_c$ is a color count, $o_{hw}$ output occupancy and $a$ one stored positive scalar. Even evaluating this actual expression exactly and rounding once predicts the wrong count-two answer (Table~\ref{tab:decisioncontrast}).

\begin{table}[h]
\caption{At an occupied location, positive output selects a color. Values are in units of the smallest binary32 subnormal, $u=2^{-149}$. The count-two error adds a nonwinning color. Conditional calculations are not a native execution trace.}
\label{tab:decisioncontrast}
\centering
\begin{tabular}{@{}crrr@{}}
\toprule
Count & Final-only rounding & Conditional stepwise & Recorded native\\
\midrule
1 & 0 & 0 & 0\\
2 & 1 & 0 & 0\\
3 & 5611 & 5611 & 5611\\
\bottomrule
\end{tabular}
\end{table}

The exact count-two polynomial lies just above $u/2$ and rounds to $u$. Under the source-derived sequential model, the final products instead round $5070u\to50u\to0$: intermediate rounding moves the last product below that boundary. In a separate author-run September precision contrast, unchanged P002 passes 265/265 complete stored targets; a derivative retaining the equation, order and scalar value but using binary64 before final binary32 conversion passes 0/265. The latter adds the predicted nonwinning colors. This is an observable answer difference, not merely an amplitude discrepancy.

The result licenses a narrower conclusion than either ``the search was wrong'' or ``rounding explains everything.'' The first successful version P001 and tuned P002 both add a scalar \emph{and} exceed the old atom bound; neither is shown to have been generated and wrongly discarded. The scalar's isolated causal effect is unknown. The conditional source model and current typed execution do not recover July's internal execution path. Appendix~\ref{app:numericaldetail} preserves the source excerpts, candidate distinctions, exact scalar and accounting; Appendices~\ref{app:schedule}--\ref{app:witness} give the arithmetic and separate diagnostic witness, and Appendix~\ref{app:precision} gives the precision protocol.

The transferable operation is to compare \emph{what the exclusion preserves} with \emph{what the evaluator observes}. Here they disagree, so the support rationale cannot prohibit the executable extension. That permits a bounded distinguishing test; it does not guarantee a new solution or justify discarding the earlier search's valid failures.

%% file: handoff.tex
\paragraph{A received distinction can still disappear.}
An explicit distinction can still be lost when results are summarized. In ROGII's C12 test-cohort lane, the coordinator initially called the proposed family genuinely novel and corpus-verified. The lane's broader audit found predecessors omitted from the graph's indexed sources and stopped before efficacy testing. The report explicitly left the proposed operator's efficacy \emph{not measured}. Its executive section was returned to the coordinator, which correctly acknowledged closure ``at its novelty gate, before any building.'' About 51 minutes later, the same coordinator summarized ``every mechanism cell measured and closed,'' explicitly including C12. The aggregate summary no longer preserved the distinction between a novelty stop and an efficacy verdict. This reporting scope loss does not establish a mistaken belief or blocked experiment: a later board update again preserved the qualification. A decision record must distinguish \emph{not novel}, \emph{not measured} and \emph{measured failure} through summary compression (Appendix~\ref{app:handoff}).

\paragraph{Three authorizations, not one score.}
These cases motivate preserving which decision a result settles through five handoff items: \emph{proposition and tested scope; program/evaluator identity; test or counterexample; gate-status vector; next action and reopening condition}. Permission to \emph{probe}, eligibility to \emph{promote}, and warrant to \emph{exclude} are different judgments. A lawful, affordable probe can resolve unknown efficacy; that unknown cannot authorize promotion or an efficacy-based exclusion. Exclusion requires counterevidence covering the proposed route, not just one prior candidate. C12's novelty failure can close a novelty-funded lane without refuting the exact operator. Figure~\ref{fig:handoff} gives a retrospective interface, not a policy fitted to the 97 cost states.

\begin{figure}[H]
\centering
\begin{tabular}{@{}p{0.95\linewidth}@{}}
\toprule
\textbf{Dispatch one proposed route under a stated contract}\\
1. Is the evidence linkage or gate applicability needed for this decision unresolved?\\
\quad Yes $\to$ \textbf{check/reconcile} it; do not convert unknown into a prohibition or a pass.\\
2. Does an established, applicable constraint rule out this route?\\
\quad Yes $\to$ \textbf{stop this route}; another component or representation remains open.\\
3. Otherwise, which uncertainty should the next bounded check resolve?\\
\quad Candidate ready, required check not yet run $\to$ \textbf{validate} the next required gate.\\
\quad Candidate incomplete or failing a required check $\to$ \textbf{repair}, with an explicit retest.\\
\quad Admissible alternative escapes only the old assumptions $\to$ \textbf{reopen}, with a distinguishing prediction.\\
\quad No concrete route yet $\to$ bounded \textbf{discovery} to formulate one.\\
\textbf{Budget guard:} if no justified, affordable next check is selected, \textbf{park} with a reopening condition; do not infer impossibility.\\
\bottomrule
\end{tabular}
\caption{Proposed case-grounded dispatch tree. Steps 1--2 take precedence for the particular route; step 3 does not rank competing opportunities. A candidate already eligible for the requested gate needs no additional research check for that decision. An unverified worker bound motivates checking or provisional redirection, not a certified stop. No threshold or expected gain is learned from the archive.}
\label{fig:handoff}
\end{figure}

\paragraph{Formalizing promotion eligibility.}
Of the three authorizations, we formalize only promotion. The dispatch tree selects the next research action; the predicate specifies the evidence required before a candidate may advance. Let $\Gamma$ denote the evaluation contract, $p$ the exact candidate, $k$ the gate and $E$ the retained evidence. The nonempty required-check set $\mathcal Q_{\Gamma,k}(p,E)$ contains predeclared gate checks, including any improvement criterion, and applicable regression inputs. Required checks with unresolved applicability remain until reconciled. Define $v_E(p,q)\in\{\mathrm{pass},\mathrm{fail},\mathrm{unknown}\}$ from results bound to that candidate, evaluator, comparator and population. Separately, $a_E(q,\Gamma,k)$ encodes supported, contradicted or unresolved applicability and adequacy using the same statuses. Neither declaration nor a passing value alone justifies applicability. Missing, mismatched or unresolved conflicting evidence is $\mathrm{unknown}$, not a demonstrated failure; an unrelated pass cannot erase a failed check. Then
\begin{equation}
\label{eq:advance}
\operatorname{eligible}_{\Gamma,k}(p,E)
=\mathbf{1}\!\left[\ \forall q\in\mathcal Q_{\Gamma,k}(p,E),\quad
\bigl(a_E(q,\Gamma,k)=\mathrm{pass}\bigr)\land\bigl(v_E(p,q)=\mathrm{pass}\bigr)\ \right].
\end{equation}
The indicator is 1 when every requirement holds and 0 otherwise: gate eligibility, not correctness beyond the contract or an instruction to deploy. Applicability requires justification linking the check to the target and legal information; notation cannot certify it. In ROGII's retained S3 adaptation experiment, all three prefix-based gates selected worse pooled target outcomes than ungated adaptation. The strictest gate was worst, although the middle threshold was not monotonically worse (Appendix~\ref{app:handoff}). This challenges that proxy in that experiment, not gating in general. Keep a vector of results: unknown blocks an affirmative promotion claim, not investigation. A changed candidate must retest still-applicable failures; old failure alone does not refute new bytes. If only a receipt survives, recovering a suitable regression is a \emph{check} action, not a completed test. The formula exposes the obligation to justify gates; it cannot make a weak evaluator sound.

\paragraph{One completed handoff: Task 188, duplicated-tile recovery.}
This is an author-reconstructed record, not a historical dispatcher input. \textbf{Proposition/scope:} the earlier cost-109 program fails the stored-case gate. \textbf{Identity/evidence:} D188-E07 identifies candidate \texttt{f1a9822b\ldots}, with 94/266 incorrect. \textbf{Gate:} stored-case failure; withhold that candidate, not every cost-109 construction. \textbf{Next action:} test a distinct controller rewrite under the required checks. \textbf{Reopening condition:} new bytes and candidate-linked evidence, not a new filename or the same cost. The later \texttt{0694c066\ldots} passes 266/266 (D188-E09); that establishes its stored-case result, not general correctness. This contrast is reconstructed with the outcome known. It does not show that the earlier failure caused the rewrite or that the proposed handoff was consumed.

The distinction transfers without carrying NeuroGolf's arithmetic into another domain: specify the proposition, check whether evidence applies, select the unresolved question, and separately check the candidate's eligibility. We do not fit success probabilities, saved-compute estimates or optimal budgets: the cost states lack observed policy actions and exposure, and the selected episodes lack a counterfactual control. A decision tree makes the proposed practice inspectable, not empirically validated. The packet supplies this interface and worked records; its checks verify evidence consistency, not the policy's effectiveness.

%% file: decision_appendix.tex
\section{Evidence behind the next-action decisions}
\label{app:decisions}

The new decision packet is a retrospective reconstruction, not a benchmark of
the proposed handoff protocol. Its four episode cards, D196, D188, D364 and
D368, distinguish the prior proposition, observable worker input, pre-outcome
reasoning, test/receipt and subsequent disposition. PCOVERAGE and PCONTRACT
describe human process requests followed by agent interpretation and
implementation. All task-level construction diagnoses are agent-led; the
operator did not personally review individual solutions. Some operator inputs
forwarded task-specific candidate material or existence hints, whose authorship
is not inferred from forwarding.

\subsection{Selection, units and what the checks reproduce}

The inventory includes all 400 tasks and all 29,591 competition ledger rows.
Coverage records exist for 318 tasks; approach ledgers cover only 29 tasks.
The retained floor-registry history has six snapshots and 755 raw key rows.
Fourteen tasks have padded and unpadded current keys, sometimes with different
costs or strike counts. The analysis preserves each key and selects the first
observed qualifying task/cost state, not the first time that an agent might
have judged the task. The resulting 97 states span 92 tasks. Repeated strikes
are not verified independent judgments, and observed Git times are not
independently authenticated historical times.

All 97 states have deadline costs linked through the frozen archive's member
identities. Sixteen have an earlier lower accepted candidate, 79 a later lower
acceptance, and their deadline costs partition into 51 lower, 35 equal and 11
higher than the recorded state. Thirty-seven end lower without an earlier
lower acceptance. The 91 exact canonical-key-eligible states have corresponding
counts 11, 74 and 45/35/11; 36 of those end lower without prior lower
acceptance. Task 095 is the difference between 37 and 36: its padded-only key
cannot match the retained ranker's unpadded lookup if those exact bytes were
used. These eligibility conditions do not establish the contemporaneous bank
cost, historical scratch-registry state, other filters or actual exclusion.

Twenty-nine states have a later lower acceptance without a lower endpoint.
Six have later negative rechecks of a lower accepted hash. A comparative gate
failure is not automatically an incorrect algorithm; non-retention alone does
not identify rollback's cause. Earlier lower acceptances likewise do not prove
that a floor is stale: they may have failed another gate or concerned a
different representation. The tables are therefore cost-state contrasts, not
semantic labels of false impossibility statements.

The detailed episodes were not sampled randomly. Tasks 196,188 and 364 were
inspected as development examples; Task 368 was selected partly because its
cohort endpoint did not improve. This is outcome-informed negative-contrast
sampling. Its explicit scoped reasoning, not its unchanged score, supports a
stopping example. The already inspected perfect-score families are also
selected cases. No precision/recall statistic or held-out policy prediction
is computed from these endpoints.

\input{decision_excerpts}

\subsection{Task 188: three batches, several programs, two decisions}

The earlier worker's ten failures concern a named factorization after operand
changes, without a hash of the tested bytes. A later verifier returns the
cost 127 artifact's hash before a fresh batch finds five failures. Those two
connections have different provenance strength. Table~\ref{tab:188batches}
keeps all three batches and their comparators separate. The paired repair
changes certificate structure and operand arrangement together; it is not a
single-variable ablation. Ambiguous duplicated grids can admit two latent
generator explanations, so reference agreement and latent-label agreement
answer different questions.

\begin{table}[h]
\caption{Task 188 (recover the original duplicated tile): separate fresh checks
in D188. The final row compares all programs on one batch; earlier rows are
different batches. Numbers are retained results, not new executions.}
\label{tab:188batches}
\centering
\begin{tabular}{@{}p{0.24\linewidth}p{0.26\linewidth}p{0.40\linewidth}@{}}
\toprule
Batch & Program linkage & Recorded result\\
\midrule
Earlier rejection & Named cost 127 path/build; exact tested hash absent &10 failures among 2000 fresh examples\\
After local promotion & Accepted cost 127 hash linked to fresh check &1995/2000 latent-label matches\\
Paired repair check & Unfactored 129 / factored 127 / repaired 115 &1999/1992/1999 latent labels; 0/7/0 differences from 129\\
\bottomrule
\end{tabular}
\end{table}

The later controller worker demonstrably reads the coverage record before
proposing its change. It treats the rank-two spatial writer as minimal in its
form and changes the controller instead. This records the worker's assessment,
not an independently established minimality theorem. The accepted 109 program differs from
an earlier 109 program that failed 94/266 stored examples. Both the negative
artifact result and the component restriction remain useful, but neither
should be applied using task number or cost alone. D188 exports separate
candidate aliases and observed versus unknown consumption rather than treating
all equal-cost programs as a common lineage.

\subsection{Contrasts that prevent success-only advice}

\paragraph{Task 196: intact-frame recoloring.}
The old record includes concrete leakage, interface and runtime failures,
an overbroad propagation-only sentence, and a sensible exception for a
genuinely different idea. A worker later reads prior approaches and predicts
that nine bounded perimeter signatures can avoid propagation before it
builds the successful program. Cost 2360 becomes 2300; stored and fresh receipts
report 266/266 and 2000/2000. The separate floor note is not observed as an input
to that worker. This refutes neither every prior
counterexample nor the conditional reopening rule, and does not establish a
causal memory intervention.

\paragraph{Task 364: L/U/H component recoloring.}
The worker first reports an exact NumPy classifier and then identifies its
ONNX realization as unresolved. The nominal 4668 prototype's builder leaves
the U class unimplemented: its logit remains negative. The bounded execution returns exit code 124.
It is not a complete cheap solution. This blocks promotion of that prototype,
not all further attempts to realize a related algebraic idea.

A different, later artifact is locally accepted at 4888. The final archive
retains 5067; quarantine and submission records attribute the rollback to a
failed singleton test. The final manifest reports control 7653.42 versus
singleton 7636.95, whereas the original singleton manifest has control 7652.87.
The capsule preserves both values. It does not authenticate the remote state
or isolate the reason for rejection. The partial prototype, later accepted
artifact and deadline incumbent are separate evidence objects.

\paragraph{Task 368: copying a prototype into gray rectangles.}
A worker reads the prior coverage/sparse-relation record, applies its recorded
10-by-10 identity-minor argument only to the static translation-tensor,
direct-contraction form, and considers alternatives. A small-state recurrence
requires excluded operators; another decoding/stamping proposal lacks an
admissible primitive or an affordable standard-layout fallback. The worker
stops those routes and retains 852. Some previously documented constraints are
reconsidered, so this is not an ideal execution or measured saving. It is a
source-linked example of the decisions an earlier applicability check could
support, without claiming global optimality.

\subsection{Operator requests and implemented process changes}

PCOVERAGE links a human request to avoid starvation, the agent's three-day
interpretation and first source change, the explicit before-deadline correction,
and the all-task-round changes that follow. The source retains a bounded
one-hypothesis prompt, normal validation gates, a concise note and a stop.
This verifies the implemented interpretation change, not exact dispatch
exposure, all-task completion or progress caused by it.

PCONTRACT links the operator's four failure categories---incremental objective,
wrong assumptions, missing knowledge and genuine difficulty---to an initial
written discovery protocol. The protocol initially requires 70 generated
configurations before 60 stored cases. A later request to identify bottlenecks
precedes the agent diagnosis. A subsequent request to connect math with
implementation precedes the revised fixed-corpus-primary contract, with broader
robustness retained as a separate annotation. The capsule does not infer that the successful
candidate had already been generated and rejected. Nor does fixed-corpus
acceptance establish generalization or authorize weakening a required
deployment constraint after observing a result.

\subsection{A copyable record, with explicit unknowns}

For a consequential experiment, retain the proposition; component or family
and assumptions; immutable program identity when known; evaluator and test
population; counterexample or receipt; furthest validation gate; next action;
and stopping/reopening condition. Identify whether the next worker actually
received the evidence, rather than equating storage with use. An unknown hash,
missing read or untested alternative remains unknown.

The anonymous \path{decisions/README.md} supplies the record and evidence
route. Run \texttt{python3 decisions/verify.py} from the supplement to check
cohort arithmetic, identity/selection fields and chronology in the exported
packet. The top-level \texttt{verify.py} includes these checks. They do not
authenticate complete private sources, execute any historical candidate or
evaluate the proposed policy. Source-holder checks separately bind the selected
excerpts and receipts to retained snapshots. Full sessions and private maps
remain unbundled; this boundary is not disguised as public reproducibility of
the original campaign.

%% file: decision_excerpts.tex
\subsection{Selected contemporary excerpts}

These excerpts are transcribed from the anonymous decision packet, with whitespace
normalized and inline code typeset in monospace. They report what the recorded
actor stated, not independently proved optimality or a reconstructed cause of
success. D188-E08's controller proposal is quoted in the body; D188-E10 links
the preceding returned coverage note. The cards retain fuller context, event
order, receipts and source-consistency boundaries.

\paragraph{D196-E01: Prior restriction; its cited failures concern earlier constructions.}
\begin{quote}
``Given (1)+(2), the ONLY legal strategy is whole-grid-uniform bounded propagation (truncated flood / bitset row propagation)''
\end{quote}

\paragraph{D196-E02: Reopening exception in the same note.}
\begin{quote}
``a genuinely new whole-grid-uniform mathematical trick (not just re-packing bits)''
\end{quote}

\paragraph{D196-E03: Agent's alternative before its construction result.}
\begin{quote}
``A simpler exact extraction emerged from the bounded generator: pack each column into a 15-bit word, identify all nine possible 3–5 by 3–5 rectangle signatures with word-level ANDs, then render their perimeters directly from the anchor words. This avoids connected-component propagation entirely; I’m validating that closed-form signature now before compiling it.''
\end{quote}

\paragraph{D188-E04: Pre-repair builder comment; certificate and operand order both change.}
\begin{quote}
``Unlike the old signed radix hash this cannot cancel between rows or colours, so it removes the colour code and row-radix factors.''
\end{quote}

\paragraph{D368-E02: Agent's scoped stop; the sub-100 estimate is not a measured program.}
\begin{quote}
``\texttt{Scan} would have produced an estimated sub-100 construction, but the scorer explicitly rejects \texttt{Scan} and \texttt{Loop}; that route is closed before implementation.''
\end{quote}

\paragraph{PCOVERAGE-E04: Human correction of the implemented interpretation.}
\begin{quote}
``It's not three-day round robin, it shoud rotate all tasks before the deadline.''
\end{quote}

\paragraph{PCONTRACT-E02: Initial agent-authored protocol, before the later successful worker.}
\begin{quote}
``4. exact full-tensor equality on all 70 finite generator outcomes; 5. all 60 stored examples and at least 5,000 fresh draws;''
\end{quote}

\paragraph{PCONTRACT-E06: Revised agent-authored target; broader correctness remains separate.}
\begin{quote}
``- A full-score hit is \texttt{60/60} under the authoritative scorer with measured cost   \texttt{<= 1}. - U70 is a secondary robustness annotation and cannot veto a fixed-60 hit.''
\end{quote}

%% file: numerical_details.tex
\section{Numerical exclusion: historical and program details}
\label{app:numericaldetail}

\subsection{What the prior report excluded}

Task 129 (family F2) fills a $3\times3$ grid with its uniquely most frequent color. It already had a cost-80 program correct on all 265 stored cases; the missing achievement was cost at most one, not recognizing the task rule. In the retained default generator and corpus, the winner occurs three times and every other color at most twice. The count profiles $(3,2,2,1,1)$ and $(3,2,1,1,1,1)$ occur 195 and 70 times. This is a threshold-separation contract, not unrestricted mode selection.

The earlier search enumerated unweighted repeated-input relations without initializers, under a practical six-atom cap and additional pruning. Importantly, its implementation already computed thresholded signatures through finite-probe float32 NumPy execution. It did not implement an exact-real semantic quotient. Separately, its returned report offered the following reason not to extend the grammar. The later explanation names the distinction at issue:

\textbf{Prior returned report, X001:} ``A one-scalar initializer does not add a new threshold relation here: it can only scale or invert the monomial, so the genuine relational support delta remains cost 0.''

\textbf{Contemporary explanation, X002:} ``counts 1--2 become exactly zero while count 3 stays positive. This simultaneously performs mode selection and the 3$\times$3 support write---precisely the composition our earlier Boolean/monotone Einsum audit could not see.''

These are contiguous source excerpts with normalized whitespace; X002 begins mid-sentence. The packet links X001 to the returned report and X002 to the explanation/candidate-write sequence (Appendix~\ref{app:mechanismtrace}). Neither excerpt is a quotation reconstructed from the present witness. The prior report also localized a near miss between selecting the mode's original positions and filling the whole output. A traceback and two negative diagnostic batches preceded the successful construction; they are preserved route failures, not a complete search census.

The first successful candidate, P001, has a linked receipt reporting 265/265 correct at cost one. A separately tuned P002 re-passes those cases and reports no failures on 20,000 custom cases; its bytes match the final archive. The later sample is not established as platform-hidden data. Both candidates exceed the earlier atom bound and add a scalar. Neither is shown to have been generated and discarded by signature deduplication. The restricted search failure remains an observation, but the support argument cannot justify dismissing the broader executable family. This does not identify scalar addition as the sole beneficial change or prove that the earlier grammar has no solution.

\subsection{The actual program: support, one rounding, and native execution}

The 681-byte P002 graph contains one \texttt{Einsum}: 23 color-count factors, one occupancy factor, and 26 references to one stored float32 scalar,
\[
a=2665831/268435456\approx0.009930994361639023.
\]
Writing $n_c$ for a color count and $o_{hw}$ for output occupancy, its idealized real expression is $o_{hw}n_c^{23}a^{26}$. Reusing the scalar does not add stored parameters. The retained helper excludes the tensors named \texttt{input} and \texttt{output} from its memory charge; this terminal single-node graph has no charged graph intermediate, and its scalar contributes one stored element. This explains cost one, not actual temporary memory or work inside the contraction. The equation alone does not specify intermediate rounding; implementation source provides a more specific model below.

Positive support is the set of locations whose entries are strictly positive. In exact arithmetic it erases the scalar distinction here: at an occupied position, every present color remains positive. It is tempting to explain the actual program by evaluating its polynomial exactly and rounding only the final result. Table~\ref{tab:three-semantics} shows why this also fails. Let $u=2^{-149}$ be the smallest positive binary32 subnormal, and let $R_{32}$ denote one nearest-even rounding with subnormals preserved.

\begin{table}[h]
\caption{P002 calculations beside historical native designated-channel measurements at occupied positions. ``Conditional stepwise'' is the source-derived recurrence below, not a native trace. Exact values are displayed approximately; the count-two comparison with $u/2$ uses exact rational arithmetic.}
\label{tab:three-semantics}
\centering
\begin{tabular}{@{}crrrr@{}}
\toprule
Count $n$ & $n^{23}a^{26}/u$ & \shortstack{Final-only\\rounding$/u$} & \shortstack{Conditional\\stepwise$/u$} & \shortstack{Recorded native\\output$/u$} \\
\midrule
1 & $0.0000000596046953$ & 0 & 0 & 0 \\
2 & $0.5000004238446835$ & 1 & 0 & 0 \\
3 & $5611.375488708714$ & 5611 & 5611 & 5611 \\
\bottomrule
\end{tabular}
\end{table}

Exactly, $n^{23}a^{26}/u=2665831^{26}/2^{556}$ at $n=2$, and $2665831^{26}>2^{555}$. The count-two value is therefore strictly above $u/2$, so a single final rounding retains that nonwinning color. Both stored count profiles contain a count-two color: the final-round calculation gives respectively 18 and 9 extra positive output entries for the profiles listed above. These are calculated consequences, not measured native mismatch counts; the discrepancy changes the task answer, not only an irrelevant amplitude.

\paragraph{Where a source-based model crosses the boundary.}
The public CPU implementation for ONNX Runtime 1.24.4 processes operands in listed order, materializing a float32 tensor after each pair~\citep{orteinsum}. Applied to P002, the independent count reductions and size-one contraction dimension give a conditional recurrence: start at $r=n$, apply $r\leftarrow R_{32}(rn)$ 22 times, multiply by occupancy, then apply $r\leftarrow R_{32}(ra)$ 26 times. For count two at an occupied location, the final two scalar products give
\[
5070u\ \xrightarrow{\ \times a,\ R_{32}\ }\ 50u
\ \xrightarrow{\ \times a,\ R_{32}\ }\ 0,
\qquad 50a=\frac{66645775}{134217728}<\frac12.
\]
Unlike the exact polynomial's value just above $u/2$, the last product is now below it. Exact rational simulation yields $(0,0,5611)u$ for counts $(1,2,3)$, matching the recorded amplitudes. No coefficient or operand order was searched in this source-model analysis. This is a post-hoc explanation under nearest-even, gradual-underflow arithmetic and the cited source order, not a traced execution: 1.24.4 is the September replay's reported version, whereas July's F2 environment is linked to 1.27.0. Neither the historical intermediates nor exact binary build/FP controls are recovered. Appendix~\ref{app:schedule} supplies the source chain, arithmetic, and falsifiers.

The historical probes measure one designated channel, not every output. An out-of-contract count-two mode is suppressed, so the construction is not a general argmax. Full-task correctness/cost receipts and present-day fixed-artifact replay are separate evidence (Appendix~\ref{app:replay}).

\paragraph{A measured precision contrast, not an intermediate trace.}
In a bounded September check, unchanged P002 passes all 265 stored cases and matches the source model bit-for-bit across all 2,385,000 output entries. A diagnostic derivative preserves its equation, operand order and scalar value but executes in binary64 before one binary32 output conversion. It passes none: count-two colors contribute the predicted 18 or 9 extra positives per case, with no missing target or positive padding entries. Thus preserving the idealized polynomial does not preserve this program's decoded answer across precisions. The typed kernels can differ; this contrast corroborates current execution dependence, not the internal rounding path, July's execution, or discovery causation. Appendix~\ref{app:precision} supplies the fixed protocol and receipt.

\subsection{A minimal counterexample isolating the support error}
\label{sec:witness}

To isolate the implication challenged by the prior support argument, consider a \emph{new mathematical witness, not the historical ONNX computation}. The coefficient $u/16$ is not representable as a binary32 initializer; this witness specifies exact arithmetic followed by one rounding, not a directly deployable one-scalar ONNX candidate. At an occupied position compare $n_c^2$ with $z_c=n_c^2u/16$, rounding the latter once by $R_{32}$:
\begin{equation}
\label{eq:witness}
\begin{array}{c|ccc}
n_c & 1 & 2 & 3 \\
z_c/u & 1/16 & 1/4 & 9/16 \\
\mathbf{1}\{R_{32}(z_c)>0\} & 0 & 0 & 1.
\end{array}
\end{equation}
Both expressions have identical positive support in exact arithmetic. After the specified rounding, only the scaled expression selects the winning color. Multiplication by $o_{hw}$ fills all nine occupied output positions and leaves padding zero, producing the complete target under the count-gap contract.

Appendix~\ref{app:witness} proves the coefficient interval, supplies negative and out-of-contract controls, and explains the finite-signature checks. The runnable witness tests all 9,000 decoded entries of a synthetic grid. It isolates a failure of exact-support equivalence; it must not be substituted for P002's different, intermediate-rounding computation.

%% file: spatial_appendix.tex
\section{Spatial support and execution-gate evidence}
\label{app:spatial}

\paragraph{Visible rotation is not the complete output.}
Family F1 includes Task 053's downward shift and four adapted recipients: 135 (upper-right crop), 326 (upper-left crop), 087 (180-degree rotation), and 140 (sparse 180-degree rotation). Their common object is task-adapted spatial support, not an unchanged universal program. The selected first rotation attempt fails all 266 stored Task 087 cases and all 265 Task 140 cases; a later construction passes each corpus. Table~\ref{tab:padding-map} gives the map on each spatial output axis recovered by static analysis of the two construction calls.

The recorded calls use per-axis $(k,d,s,p_{\mathrm{begin}},p_{\mathrm{end}})$ values $(2,3,-1,-2,-53)$ and $(3,10,-11,-2,-326)$, where $k$ is kernel size, $d$ dilation, $s$ stride and $p$ padding. Their source-index windows are
\[
W_i=\{is-p_{\mathrm{begin}}+jd:j=0,\ldots,k-1\}.
\]
The failed form gives $\{2-i,5-i\}$; the later form gives $\{2-11i,12-11i,22-11i\}$. Intersecting each with logical indices $\{0,1,2\}$ derives Table~\ref{tab:padding-map}. The signed attributes are recorded runtime-specific choices, not a standard positive-stride pooling identity. The supplement discloses the literal attributes and declared interface in \path{replay/F1_CONSTRUCTION.json}; its diagnostic recomputes the maps from those attributes.

\begin{table}[h]
\caption{Source-coordinate reconstruction of the selected rotation pair. Both implement the visible half-turn; the failed form additionally repeats logical-input support outside the visible grid. ``None'' means no selected logical-input support, not a newly measured runtime value.}
\label{tab:padding-map}
\centering
\begin{tabular}{@{}lll@{}}
\toprule
Output positions & Failed form & Later form \\
\midrule
$0,1,2$ (visible) & Input $2,1,0$ & Input $2,1,0$ \\
$3,4,5$ (padding) & Input $2,1,0$ & None \\
Remaining padding & None & None \\
\bottomrule
\end{tabular}
\end{table}

For a synthetic grid with rows $(1,2,3)$, $(4,5,6)$, $(7,8,9)$, the failed separable map puts four copies of the rotated grid in a $6\times6$ region. It has 27 extra positive entries outside the target; the corrected map has none. Both match the visible nine cells. The shipped \texttt{diagnostics.py} compares all 9,000 entries and also checks the disclosed P003 coordinate law below. These are calculated support-model outputs, not measured mismatch counts from the historical run.

The changed property is not recognition of the rotation rule. It is support outside the logical $3\times3$ grid, which the strict full-tensor decoder also checks. Matching visible output therefore does not establish the required equality. The source-coordinate reconstruction is explanatory evidence, not a newly measured runtime trace or a controlled padding-only ablation; the aggregate failures do not localize every actual mismatch. The early in-memory programs have unrecovered byte identities. Later hashed validation, promotions, shared final bytes for 087/140, and replay are linked separately rather than silently identifying every passing construction with the archived program.

The separately byte-identified archived P003 has a more explicit coordinate account. The recorded \texttt{MaxPool} construction pools each color channel over windows whose candidate input indices are $W_i=\{2-11i,12-11i,22-11i\}$ on each output axis. In-canvas positions outside the logical grid are zero-encoded. Intersecting with the logical input positions yields
\[
W_i\cap\{0,1,2\}=\begin{cases}\{2-i\},&0\leq i\leq2,\\ \emptyset,&3\leq i\leq29.\end{cases}
\]
Thus the recorded law reverses each visible axis while selecting no logical input in padding. Applying it on both axes gives the full-canvas half-turn under the recorded support interpretation. This explains the successful map without identifying P003 with the earlier repaired bytes. The coordinate argument is not a portable operator-semantics or native cost proof; those remain separate execution claims.

\paragraph{Execution is a prerequisite for observing correctness.}
Task 053 supplies a different inferential boundary. Among 105 distinct primary candidates in nine retained late cycles, 53 are rejected at runtime/session initialization and one at the interface. Of the 51 that execute, 12 are correct on all 60 stored cases but exceed cost one; 39 meet the cost cap but fail all 60. The target is joint membership in execution, tested correctness, and cost eligibility, $E\cap C\cap B$. Separate successes in $C$ and $B$ do not imply an accepted candidate. Rejected executions have unestablished semantics, not measured zero accuracy.

Only 92 of these primary candidates returned before the successful worker's dispatch: 47 rejects, 12 correct over-cost candidates, and 33 cost-eligible wrong candidates. The last cycle overlaps that investigation. The worker's distinct preflight batches first yield 33 shape/dimension checker rejections, then 24 checker-compatible records with 22 executions and two session failures. A later full-task batch has 47 configurations passing 60/60 at cost zero and 41 checker rejects. These are separate batches, not a controlled intervention or additions to the 105-candidate denominator. They establish what each gate observed, not why a worker changed its approach.

The subsequent integration cycle records a provisional pivot with runtime, full-output, inactive-region, and falsifier conditions. This is post-success consolidation, not evidence that memory originated the construction. Further selected branches remain negative: Task 385's mirror-completion adaptation passes 0/265, and broader fixed-artifact screens add no task. These outcomes bound the examined attempts, not entire tasks or mechanism families; complete screen denominators remain in Appendix~\ref{app:screenbounds}.

%% file: handoff_appendix.tex
\subsection{Receipt, summary and gate scope}
\label{app:handoff}

Two additional ROGII episodes challenge the interface in Section~\ref{sec:broad}. They were selected after the NeuroGolf analysis, from the same operator's earlier campaign; they are neither a held-out evaluation nor observations of our formalized protocol in use. The new \path{cross_domain/HANDOFF.md} route is separate from the augmentation/input-scope comparison above. It packages selected visible records and aggregate measurements, not private conversations, well identifiers or executable models.

\paragraph{C12: novelty closure is not an efficacy result.}
The proposed test-cohort operator would allow hidden wells to constrain one another at inference. The lane required a novelty audit before implementation. A no-hit in the knowledge graph and its indexed digests did not cover the whole project: the manifest omitted relevant local experiments and public writeups. Expanded retrieval found prior work on this family. The report therefore rejected the broad novelty premise, while preserving the exact proposed 50/100-ft operator as unmeasured and potentially a narrower ablation. Its target improvement threshold was not evaluated. This is not evidence that the exact operator could never help, nor that it should have been funded after the novelty premise failed.

\begin{table}[H]
\caption{Selected C12 chronology, relative to the initial coordinator claim. These are visible recorded statements, not inferred thoughts. The report return and its subsequent acknowledgment establish a delivery link; the later broad summary does not establish a downstream exclusion.}
\label{tab:c12}
\centering
\begin{tabular}{@{}p{0.16\linewidth}p{0.76\linewidth}@{}}
\toprule
Elapsed & Recorded event and justified scope\\
\midrule
0 & Coordinator calls the proposed family genuinely novel.\\
20 min 10 s & Report executive section returned: efficacy fields remain \emph{not measured}; novelty gate is NO-GO.\\
20 min 22 s & Coordinator correctly acknowledges the novelty gate and no building.\\
71 min 29 s & Coordinator says every mechanism cell was measured and closed, explicitly including C12.\\
13 h 8 min 45 s & Later board summary again preserves C12's novelty qualification.\\
\bottomrule
\end{tabular}
\end{table}

The acknowledgment-to-summary interval is 3067.545 seconds. This supports a narrow evidence-consumption observation: an explicitly received distinction was initially preserved, then lost in one broader summary. It does not show persistent corruption of the knowledge graph; a later board update again records the novelty qualification. Nor does it prove that a future candidate was suppressed, that the exact operator was ever efficacy-tested in this lane, or that enforcing the proposed record would improve the final score. The source-holder check matches selected visible excerpts and call/return linkage without using hidden reasoning. Public checks establish the packet's chronology and consistency, not independent authenticity of the private session.

\paragraph{S3: the gate can be the questionable assumption.}
In a separate retained experiment, relative prefix-tail RMSE improvement was used to accept per-well test-time adaptations before evaluating the actual target zone. This gate used legally available prefix labels, not target-zone labels. Table~\ref{tab:s3} retains every reported threshold for the primary setting: one checkpoint, fold and adaptation dose, 155 wells and 750,596 target rows. The matched baseline is 6.5193, not an older prediction bank's 6.5187. Lower pooled RMSE is better; each gated arm uses the adapted result on accepted wells and the baseline elsewhere.

\begin{table}[H]
\caption{Retained S3 pooled target RMSE (ft), not a new model run or mean per-well RMSE. All gated arms are worse than ungated adaptation and the matched baseline; the threshold sequence is not monotonic.}
\label{tab:s3}
\centering
\begin{tabular}{@{}lrr@{}}
\toprule
Arm / required prefix-tail gain & Accepted adaptations & Target RMSE\\
\midrule
Matched baseline & --- & 6.5193\\
Ungated adaptation & 155 & 6.5411\\
Gate: $0\%$ & 96 & 6.5556\\
Gate: $2\%$ & 63 & 6.5538\\
Gate: $5\%$ & 25 & 6.5718\\
\bottomrule
\end{tabular}
\end{table}

The report records correlation $-0.115$ and sign agreement 0.490 between prefix-tail and target-zone gains. These results undermine this surrogate as a reliable selection signal in the retained setting; they do not establish that all stricter gates are harmful or that stronger adaptation doses were tested. The distinction is between passing a proxy and satisfying the claimed target criterion, not between our proposed promotion rule and a control policy. The packet checks aggregate arithmetic; the private source-holder check additionally recomputes gate counts and these descriptive associations from retained per-well summaries, without training or scoring raw predictions. Gate applicability remains a scientific judgment requiring evidence, not a Boolean made true by preregistration.

%% file: main.bbl
\begin{thebibliography}{14}
\providecommand{\natexlab}[1]{#1}
\providecommand{\url}[1]{\texttt{#1}}
\expandafter\ifx\csname urlstyle\endcsname\relax
  \providecommand{\doi}[1]{doi: #1}\else
  \providecommand{\doi}{doi: \begingroup \urlstyle{rm}\Url}\fi

\bibitem[Chollet(2019)]{arc}
Fran{\c{c}}ois Chollet.
\newblock On the measure of intelligence.
\newblock \emph{arXiv preprint arXiv:1911.01547}, 2019.
\newblock URL \url{https://arxiv.org/abs/1911.01547}.

\bibitem[{CroDoc} and {Pavel}(2026)]{neurogolf8}
{CroDoc} and {Pavel}.
\newblock 8th place gold---pavel \& crodoc.
\newblock NeuroGolf competition writeup, Kaggle, 2026.
\newblock URL
  \url{https://www.kaggle.com/competitions/neurogolf-2026/writeups/8th-place-gold-pavel-and-crodoc}.
\newblock Author-reported experience; accessed September 9, 2026.

\bibitem[de~Andrade(2026)]{neurogolf1pipeline}
Eduardo~Rocha de~Andrade.
\newblock 1st place---kaggle agent---pipeline overview.
\newblock Kaggle discussion, 2026.
\newblock URL
  \url{https://www.kaggle.com/competitions/neurogolf-2026/discussion/726799}.
\newblock Author-reported experience; accessed September 9, 2026.

\bibitem[{franksunp}(2026)]{initialbaseline}
{franksunp}.
\newblock {[7245.63 LB] NeuroGolf Variant B Mark A}.
\newblock Kaggle notebook, 2026.
\newblock URL
  \url{https://www.kaggle.com/code/franksunp/7245-63-lb-neurogolf-variant-b-mark-a}.
\newblock Current page title; accessed September 8, 2026. Retained
  starting-archive metadata targets version 58 and reports 7240.26.

\bibitem[Gozeten et~al.(2026)Gozeten, Zhang, Ildiz, Taga, Javidi, and
  Oymak]{emosta}
Halil~Alperen Gozeten, Xuechen Zhang, Emrullah Ildiz, Ege~Onur Taga, Tara
  Javidi, and Samet Oymak.
\newblock Evolutionary multi-task optimization for {LLM}-guided program
  discovery.
\newblock \emph{arXiv preprint arXiv:2605.22613}, 2026.
\newblock URL \url{https://arxiv.org/abs/2605.22613}.

\bibitem[{hongan} et~al.(2026){hongan}, {MaxChen303}, Sypetkowski,
  Ja{\'s}kowska, and {mrmldjr}]{neurogolf4}
{hongan}, {MaxChen303}, Maciej Sypetkowski, Natalia Ja{\'s}kowska, and
  {mrmldjr}.
\newblock 4th place solution writeup.
\newblock NeuroGolf competition writeup, Kaggle, 2026.
\newblock URL
  \url{https://www.kaggle.com/competitions/neurogolf-2026/writeups/4th-place-solution-writeup}.
\newblock Author-reported experience; accessed September 8, 2026.

\bibitem[{Kaggle}(2026)]{neurogolf}
{Kaggle}.
\newblock The 2026 {NeuroGolf} championship, 2026.
\newblock URL \url{https://www.kaggle.com/competitions/neurogolf-2026}.

\bibitem[Kapoor et~al.(2024)Kapoor, Stroebl, Siegel, Nadgir, and
  Narayanan]{agentsmatter}
Sayash Kapoor, Benedikt Stroebl, Zachary~S. Siegel, Nitya Nadgir, and Arvind
  Narayanan.
\newblock {AI} agents that matter.
\newblock \emph{arXiv preprint arXiv:2407.01502}, 2024.
\newblock URL \url{https://arxiv.org/abs/2407.01502}.

\bibitem[Liu et~al.(2025)Liu, Zhu, Chen, and Jiang]{deepevolve}
Gang Liu, Yihan Zhu, Jie Chen, and Meng Jiang.
\newblock Scientific algorithm discovery by augmenting {AlphaEvolve} with deep
  research.
\newblock \emph{arXiv preprint arXiv:2510.06056}, 2025.
\newblock URL \url{https://arxiv.org/abs/2510.06056}.

\bibitem[{Microsoft}(2026)]{orteinsum}
{Microsoft}.
\newblock {ONNX Runtime} 1.24.4: {CPU Einsum} implementation.
\newblock Implementation source, 2026.
\newblock URL
  \url{https://github.com/microsoft/onnxruntime/blob/2d924974ef147392ced8409d36bd6d2e7fcc8a74/onnxruntime/core/providers/cpu/math/einsum_utils/einsum_typed_compute_processor.cc}.
\newblock Immutable source revision 2d924974ef147392ced8409d36bd6d2e7fcc8a74;
  accessed September 9, 2026.

\bibitem[Moffitt(2025)]{arcgen}
Michael~D. Moffitt.
\newblock {ARC-GEN}: A mimetic procedural benchmark generator for the
  abstraction and reasoning corpus.
\newblock \emph{arXiv preprint arXiv:2511.00162}, 2025.
\newblock URL \url{https://arxiv.org/abs/2511.00162}.

\bibitem[{robga}(2026)]{representation}
{robga}.
\newblock 21st place solution---representation collapse.
\newblock NeuroGolf competition writeup, Kaggle, 2026.
\newblock URL
  \url{https://www.kaggle.com/competitions/neurogolf-2026/writeups/21st-place-solution-representation-collapse}.
\newblock Cited by title; current page rank differs. Author-reported
  experience; accessed September 8, 2026.

\bibitem[Shinn et~al.(2023)Shinn, Cassano, Berman, Gopinath, Narasimhan, and
  Yao]{reflexion}
Noah Shinn, Federico Cassano, Edward Berman, Ashwin Gopinath, Karthik
  Narasimhan, and Shunyu Yao.
\newblock Reflexion: Language agents with verbal reinforcement learning.
\newblock \emph{arXiv preprint arXiv:2303.11366}, 2023.
\newblock URL \url{https://arxiv.org/abs/2303.11366}.

\bibitem[Wang(2026)]{neurogolf9}
Yiheng Wang.
\newblock 9th place solution.
\newblock NeuroGolf competition writeup, Kaggle, 2026.
\newblock URL
  \url{https://www.kaggle.com/competitions/neurogolf-2026/writeups/9th-place-solution}.
\newblock Author-reported experience; accessed September 8, 2026.

\end{thebibliography}
